\documentclass[12pt]{article}
\usepackage{epsfig,cite,lscape}

\input paperdef

\begin{document}

\thispagestyle{empty}
\setcounter{page}{0}
\def\thefootnote{\fnsymbol{footnote}}

\begin{flushright}
CALT-68-2407, DCPT/03/02\\
IPPP/03/01, LMU 18/02\\
TUM-HEP-481/02 \\
hep-ph/0302174 \\
\end{flushright}

\vspace{1mm}

\begin{center}

{\large\sc {\bf The lightest Higgs Boson of mSUGRA, mGMSB and mAMSB}}

\vspace{0.4cm}

{\large\sc {\bf at Present and Future Colliders:}}

\vspace{0.4cm}

{\large\sc {\bf Observability and Precision Analyses}}
 
\vspace{1cm}

{\sc 
A.~Dedes$^{1}$%
\footnote{email: dedes@ph.tum.de}%
, S.~Heinemeyer$^{2}$%
\footnote{email: Sven.Heinemeyer@physik.uni-muenchen.de}%
, S.~Su$^{3}$%
\footnote{email: shufang@theory.caltech.edu}%
~and G.~Weiglein$^{4}$%
\footnote{email: Georg.Weiglein@durham.ac.uk}
}

\vspace*{1cm}

{\sl
$^1$ Physik Department, Technische Universit\"at M\"unchen,
D-85748 Garching, Germany

\vspace*{0.4cm}

$^2$Institut f\"ur theoretische Elementarteilchenphysik,
LMU M\"unchen, Theresienstr.\ 37, D-80333 M\"unchen, Germany

\vspace*{0.4cm}

$^3$California Institute of Technology, Pasadena, CA 91125, USA

\vspace*{0.4cm}

$^4$Institute for Particle Physics Phenomenology, University of Durham,\\
Durham DH1~3LE, UK

}

\end{center}

\vspace*{0.2cm}

\begin{abstract}
We investigate the physics of the lightest $\cp$-even MSSM Higgs boson at
the Tevatron, the LHC, a linear \epem\ collider, a $\ga\ga$~collider
and a $\mu^+\mu^-$~collider. The analysis is performed in the three
most prominent soft SUSY-breaking scenarios, mSUGRA, mGMSB and mAMSB.
For all colliders the observability and parameter regions with
suppressed production cross sections (compared to a SM Higgs boson
with the same mass) are investigated. For the lepton
and photon colliders the potential is analyzed of precision measurements 
of the branching ratios of the light $\cp$-even Higgs boson for obtaining 
indirect bounds on the mass of the $\cp$-odd Higgs boson and the
high-energy parameters of the soft SUSY-breaking scenarios. In regions
of the parameter space where the LHC can detect the heavy Higgs bosons,
precision measurements of the properties of the light Higgs boson at the
linear collider can provide valuable information for distinguishing
between the mSUGRA, mGMSB and mAMSB scenarios.
\end{abstract}
%\pacs{}

\def\thefootnote{\arabic{footnote}}
\setcounter{page}{0}
\setcounter{footnote}{0}

\newpage

%%%%%%%%%%%%%%%%%%%%%%%%%%%%%%%%%%%%%%%%%%%%%%%%%%%%%%%%%%%%%%
%%%%%%%%%%%%%%%%%%%%%%%%%%%%%%%%%%%%%%%%%%%%%%%%%%%%%%%%%%%%%%

\section{Introduction}

The search for the light neutral Higgs boson is a crucial test of
Supersymmetry (SUSY) that can be performed with the present and the next
generation of high-energy colliders. The prediction of a relatively light
Higgs boson is common to all supersymmetric models whose couplings
remain in the perturbative regime up to a very high energy
scale~\cite{susylighthiggs}.  Finding the Higgs boson and subsequently
studying its couplings to fermions and bosons is thus one of
the main goals of high-energy physics.
The data taken during the final year of LEP running at $\sqrt{s} \gsim
206 \gev$ have established a 95\% C.L.\ exclusion limit for the Standard 
Model (SM) Higgs boson of $\MHSM > 114.4 \gev$. They showed a slight
excess at about the $2\sigma$ level of signal-like events over the background 
expectation, which would be compatible with the 
expectation for the production of a Higgs boson with SM-like $ZZH$ coupling
with a mass $\MHSM \approx 116\pm1 \gev$~\cite{LEPHiggs}.
In the Minimal Supersymmetric Standard Model (MSSM)
the mass of the lightest $\cp$-even Higgs boson, $\mh$, is bounded 
from above by $\mh \lsim 135 \gev$~\cite{mhiggslong,mhiggsAEC} 
(taking into account 
radiative corrections up to \twol\
order~\cite{mhiggslong,mhiggsAEC,mhiggsRG1a,mhiggsRG1b,mhiggsRG2,mhiggsEP1,mhiggsEP2,mhiggsletter,maulpaul,maulpaul2,mhiggsEP3,mhiggsEP2L}).

In the MSSM no specific assumptions are made about the underlying
Supersymmetry- (SUSY)-breaking mechanism, and a parameterization of all 
possible SUSY-breaking terms is used. This gives rise to the huge number of 
more than 100 new parameters in addition to the SM, 
which in principle can be chosen
independently of each other. A phenomenological analysis of this model
in full generality would clearly be very involved, and one usually
restricts to certain benchmark scenarios, see e.g.\
\citeres{benchmark,benchmark2,sps}.
On the other hand, models in which all the low-energy parameters are
determined in terms of a few parameters at the Grand Unification
scale (or another high-energy scale), 
employing a specific soft SUSY-breaking scenario, are
much more predictive. The most prominent scenarios in the literature
are minimal Supergravity (mSUGRA)~\cite{Hall,mSUGRArev}, 
minimal Gauge Mediated SUSY Breaking (mGMSB)~\cite{GR-GMSB} 
and minimal Anomaly Mediated SUSY Breaking 
(mAMSB)~\cite{lr,giudice,wells}. 
Analyses of the Higgs sector in these scenarios, mostly focusing only
on the maximum value of $\mh$,  have been
performed in \citeres{higgsmsugra2,higgsgmsb2,higgsamsb,AD,higgsmsugra,higgsgmsb,higgsmsugra3,DjouadiDK}.
A detailed comparison of the three soft SUSY-breaking scenarios in
terms of exclusion regions in the $\MA-\tb$-plane (where $A$ is the
$\cp$-odd Higgs boson and $\tb$ the ratio of the vacuum expectation
values of the two Higgs doublets), their compatibility
with the slight excess observed at LEP, and their corresponding SUSY particle
spectra can be found in \citere{asbs}.

In the present paper the work of \citere{asbs}
is extended to an analysis of the lightest $\cp$-even Higgs boson
phenomenology at present and future colliders.
We relate the input from the three soft SUSY-breaking
scenarios in a uniform way to the
predictions for the low-energy phenomenology in the Higgs sector,
allowing thus a direct comparison of the predictions arising from the
different scenarios. 
The high-energy parameters given in the three
scenarios are related to the low-energy SUSY parameters via 
renormalization group (RG) running, taking into account contributions up
to two-loop order~\cite{rge2l} (for a recent comparison of different
codes and 
current accuracies, see \citere{ASBScodecomp}). After transforming the
parameters obtained in this way into the corresponding
on-shell parameters~\cite{bse,FDRG1,maulpaul3}, they are used as input
for the program \fh~\cite{feynhiggs,feynhiggs1.2,feynhiggs2}.
As a result the Higgs boson mass spectrum and the Higgs decay rates
and branching ratios~\cite{hff,hdecay} have been obtained.
Further restrictions
such as from precision observables and the non-observation
of SUSY particles at LEP and the Tevatron are also taken into account.
For an analysis within the mSUGRA scenario where also the cold dark
matter (CDM) constraints are included, see \citere{ehow}.
Based on these predictions for the Higgs sector phenomenology, we
analyze the observability of the lightest MSSM Higgs boson at the
Tevatron, the LHC, a future \epem\ linear collider (LC), a 
$\ga\ga$ collider (\gaC), and a $\mu^+\mu^-$
collider (\muC). Regions of the high-energy parameter space with
strongly suppressed Higgs production cross sections are identified.
As the next step the branching ratios of the lightest Higgs boson into
SM fermions ($\hbb, \hcc, \htautau$) and into $W$~bosons ($\hWW$) are
compared for the mSUGRA, mGMSB and mAMSB case. 
We show that the precise measurement of the various Higgs decay
branching ratios can be used to impose bounds on the value of $\MA$ and
also the high-energy input parameters. 
Our analysis considerably differs from existing studies of Higgs boson
branching ratios in the literature~\cite{MAdet1,MAdet2,MAdet3}.
In these previous analyses, all parameters except for the
one under investigation (e.g. $\MA$) have been kept fixed and the effect of an
assumed deviation between the MSSM and the SM has solely been attributed to
this single free parameter. This would correspond to a situation with 
a complete knowledge of all other SUSY parameters without any experimental or
theoretical uncertainty, which obviously leads to an unrealistic
enhancement of the sensitivity to the investigated parameter. In our
analysis we performed a more realistic study allowing all the SUSY
parameters to vary. Furthermore, 
combined with the information on
$\MA$ that could be obtained from the LHC Higgs searches, we 
discuss the possibility of distinguishing the three scenarios via the
Higgs branching ratio measurements at the LC.

The rest of the paper is organized as follows. In Sect.~2 we briefly
review the three soft SUSY-breaking scenarios and the evaluation of
the low-energy data.
The observability of the lightest MSSM Higgs boson is investigated in
Sect.~3. In Sect.~4 the potential is analyzed 
of precision measurements of the Higgs boson branching ratios for
obtaining indirect constraints on $\MA$ and the high-energy parameters
of the soft SUSY-breaking scenarios. The  possibility of a distinction
of the mSUGRA, mGMSB and mAMSB scenarios is discussed.
The conclusions can be found in Sect.~5.

%%%%%%%%%%%%%%%%%%%%%%%%%%%%%%%%%%%%%%%%%%%%%%%%%%%%%%%%%%%%%%
%%%%%%%%%%%%%%%%%%%%%%%%%%%%%%%%%%%%%%%%%%%%%%%%%%%%%%%%%%%%%%

\section{The low-energy sector and phenomenological constraints}
\label{sec:higgs}

In deriving the low-energy parameters in the three soft SUSY-breaking 
scenarios (mSUGRA, mGMSB and mAMSB) from the high-energy input
parameters we follow \citere{asbs}.
Thus, in this section only the most relevant facts are briefly summarized.

%%%%%%%%%%%%%%%%%%%%%%%%%%%%%%%%%%%%%%%%%%%%%%%%%%%%%%%%%%%%%%
%%%%%%%%%%%%%%%%%%%%%%%%%%%%%%%%%%%%%%%%%%%%%%%%%%%%%%%%%%%%%%

\subsection{The soft SUSY-breaking scenarios}
\label{subsec:susybreak}

The fact that no SUSY partners of the SM particles have so far been
observed means that low-energy SUSY cannot be realized as an unbroken
symmetry in nature, and SUSY models thus have to incorporate
additional Supersymmetry breaking interactions. 
This is achieved by adding to the Lagrangian (defined by the 
${\rm SU(3)}_C\times {\rm SU(2)}_L \times  {\rm U(1)}_Y$ gauge
symmetry and the superpotential $W$)
some further interaction terms that respect the gauge symmetry but break 
Supersymmetry (softly, i.e.\ no quadratic divergences appear), so
called ``soft SUSY-breaking'' (SSB) terms.
Assuming that the $R$-parity symmetry~\cite{herbi}
is conserved, which we do in this paper
for all SUSY breaking scenarios, reduces the amount of 
new soft terms allowed in the Lagrangian.
Choosing a particular soft SUSY-breaking pattern allows further
reduction of the number of free parameters and the construction
of predictive models. The three most prominent scenarios for such
models are

\begin{itemize}

\item 
{\bf mSUGRA} (minimal Super Gravity scenario)~\cite{Hall,mSUGRArev}:\\
Apart from the SM parameters (for the experimental values of the SM 
input parameters we use~\citere{pdg}), 
4~parameters and a sign are required to define the mSUGRA scenario:
\BE
\{\; m_0\;,\;m_{1/2}\;,\;A_0\;,\;\tb \;,\; {\rm sign}(\mu)\; \} \;.
\label{mSUGRAparams}
\EE
While $m_0$, $m_{1/2}$ and $A_0$ define the scalar and fermionic
masses and the trilinear couplings at the GUT scale 
($\sim 10^{16} \gev$), $\tb$ (the ratio of the two vacuum expectation
values) and the sign($\mu$) ($\mu$ is the supersymmetric Higgs mass
parameter) are defined at the low-energy scale. 
For our numerical analysis, see \refses{sec:higgsobs},
\ref{sec:higgsbrs}, we have scanned 
over the following parameter space%
\footnote{
The sign of $\mu$ has been fixed to (+) (for all three soft
SUSY-breaking scenarios), since this sign is favored by the 
$g_\mu - 2$~\cite{gminus2} and the 
$\br(b \to s \ga)$~\cite{bsgammaexp} constraints, see
\refse{subsec:constraints}. 
}%
:
\BEA
50~{\rm GeV} \le &m_0& \le 1~{\rm TeV} \;, \nonumber \\
50~{\rm GeV} \le  &m_{1/2}& \le 1~{\rm TeV} \;, \nonumber \\
-3~{\rm TeV} \le &A_0& \le 3~{\rm TeV} \;, \non \\
1.5 \le &\tan\beta & \le 60 \;, \non \\
 &{\rm sign}\, \mu& = +1 .
\label{msugraparam}
\EEA

The low-energy spectrum has been evaluated with the programs 
{\em SUITY/FeynSSG}~\cite{sakis2,FeynSSG}.

\item 
{\bf mGMSB} (minimal Gauge Mediated SUSY-Breaking)~\cite{GR-GMSB}:\\
A very promising alternative to mSUGRA is based on the hypothesis
that the soft SUSY-breaking occurs at relatively low energy scales 
and is mediated mainly by gauge interactions through the so-called
``messenger sector''~\cite{oldGMSB,newGMSB,GR-GMSB}. 
Also in this scenario, the low-energy parameters depend on
4~parameters and a sign,
\BE
\{\; M_{\rm mess}, \; N_{\rm mess}, \; \Lambda, \; 
     \tb, \; {\rm sign}(\mu) \; \} \;,
\label{eq:pars}
\EE
where $M_{\rm mess}$ is the overall messenger mass scale; 
$N_{\rm mess}$ is a number called the 
messenger index, parameterizing the structure of the messenger
sector; $\Lambda$ is the universal soft SUSY-breaking mass scale felt by the
low-energy sector.
The phenomenology of mGMSB is characterized by the presence of a very
light gravitino  $\tilde{G}$ with mass given by  
$m_{3/2} = m_{\tilde{G}} = \frac{F}{\sqrt{3}M'_P} \simeq 
\left(\frac{\sqrt{F}}{100 \tev}\right)^2 2.37 \; {\rm eV}$~\cite{Fayet},  
where $\sqrt{F}$ is the fundamental scale of SSB and 
$M'_P = 2.44 \times 10^{18} \gev$ is the reduced Planck mass.
Since $\sqrt{F}$ is typically of order 100 TeV, the $\tilde{G}$ is always the 
LSP in these theories. 
The numerical analysis in \refses{sec:higgsobs}, \ref{sec:higgsbrs} is
based on the following scatter ranges:
\BEA
10^4 \gev \le &\La& \le 2\,\times\,10^5 \gev \;, \non \\
1.01\,\La \le &M_{\rm mess}& \le 10^5\,\La \;, \nonumber \\
1 \le  &N_{\rm mess}& \le 8 \;, \nonumber \\
1.5 \le &\tan\beta & \le 60 \;, \non \\
 &{\rm sign}\, \mu& = +1 .
\label{gmsbparam}
\EEA

The low-energy parameter sets for this scenario have been calculated
by using the program 
{\em SUSYFIRE}~\cite{SUSYFIRE} and adopting the phenomenological 
approach of \citeres{AKM-LEP2,AB-LC,AMPPR, higgsgmsb}.

\item
{\bf mAMSB} (minimal Anomaly Mediated SUSY-Breaking)~\cite{lr,giudice,wells}:\\
In this model, SUSY breaking happens on a separate brane and is  
communicated to the visible world via the super-Weyl anomaly. 
The particle spectrum is determined by 3~parameters and a sign:  
\BE
\{m_{\rm aux},\ m_{0},\ \tb,\ {\rm sign}(\mu) \} .
\label{amsbparams}
\EE
The overall scale of SUSY particle masses is set by $m_{\rm aux}$, 
which is the VEV of the auxiliary field in the supergravity multiplet.
$m_0$ is introduced as a phenomenological parameter 
to avoid negative slepton mass squares, for other
approaches to this problem see \citeres{lr,negative,clm,kss,jjw}.
The scatter parameter space for the numerical analysis in
\refses{sec:higgsobs}, \ref{sec:higgsbrs} is chosen to be
\BEA
20 \tev \le &m_{\rm aux}& \le 100 \tev , \non \\
0 \le &m_0& \le 2 \tev , \non \\
1.5 \le &\tb& \le 60 , \non \\
 &{\rm sign}\, \mu& = +1 . 
\label{amsbparam}
\EEA

The low-energy spectrum has been derived with the code described in
\citere{higgsamsb}.

\end{itemize}

%%%%%%%%%%%%%%%%%%%%%%%%%%%%%%%%%%%%%%%%%%%%%%%%%%%%%%%%%%%%%%
%%%%%%%%%%%%%%%%%%%%%%%%%%%%%%%%%%%%%%%%%%%%%%%%%%%%%%%%%%%%%%

\subsection{Evaluation of predictions in the Higgs boson sector 
of the MSSM}
\label{subsec:mssmhiggs}

The most relevant parameters for Higgs boson phenomenology in the MSSM
are the mass of the $\cp$-odd Higgs boson, $\MA$, the ratio of the two
vacuum expectation values, $\tb$, the scalar top masses and mixing
angle, $\mste, \mstz, \tst$, for large $\tb$ also the scalar
bottom masses and mixing angle, $\msbe, \msbz, \tsb$, the supersymmetric 
Higgs mass parameter, $\mu$, the U(1) and SU(2) gaugino masses,
$M_1$ and $M_2$, and the gluino mass, $\mgl$. 
These low-energy parameters are derived from the high-energy parameters
of the three soft SUSY-breaking scenarios via RG running, see
\citere{asbs}. Since the RG running employed in the three scenarios
is based on the \drbar\ scheme, the corresponding low-energy parameters
are \drbar\ parameters. In order to derive predictions for observables,
i.e.\ particle masses and mixing angles, these parameters in general have 
to be converted into on-shell parameters.

For the predictions in the MSSM Higgs sector we use results obtained 
in the Feynman-diagrammatic (FD) approach within the on-shell
renormalization scheme as incorporated in the Fortran code 
\fh~\cite{feynhiggs,feynhiggs1.2,feynhiggs2} based on
\citeres{mhiggsletter,mhiggsAEC,mhiggslong}.

Our analysis is concerned with the main Higgs production and decay
channels at different colliders. To this end the
predictions for the Higgs boson masses and effective
couplings (especially the effective mixing angle, $\aeff$,
in the neutral $\cp$-even Higgs boson sector that includes
higher-order corrections) as well as for the branching ratios of the
lightest MSSM Higgs boson (and for a SM Higgs boson with the same
mass) have been evaluated. 
The effective mixing angle $\aeff$ is defined via
\BE
 \aeff = {\rm arctan}\KKL 
  \frac{-(\MA^2 + \MZ^2) \Sb \Cb - \hSi_{\PePz}}
       {\MZ^2 \CQb + \MA^2 \SQb - \hSi_{\Pe} - \mh^2} \KKR~,~~
  -\frac{\pi}{2} < \aeff < \frac{\pi}{2}~.
\label{aeff}
\EE
Here $\hSi_s, s = \Pe, \Pz, \Pe\Pz$ denotes the renormalized Higgs
boson self-energies in the $\Pe-\Pz$ basis.
While the predictions for the decays of 
$h \to b \bar b, c \bar c, \tau^+ \tau^-, \mu^+ \mu^-$ 
are based on \citere{hff},
including the higher-order corrections described in
\citere{deltamb1,deltamb2}, the other decay channels have been derived
with the code {\em Hdecay}~\cite{hdecay}, 
which has been implemented as a subroutine
in the latest version of \fh~\cite{feynhiggs2}. The proper transition
from on-shell parameters in \fh\ to \drbar\ parameters in {\em Hdecay}
has been taken into account~\cite{bse,FDRG1,maulpaul3}.

In order to derive the relative difference between a MSSM production
or decay rate and the corresponding SM rate (for the same Higgs boson
mass), the following ratios have been calculated (below the notation 
$t\bar t \to t \bar t h$ refers to the processes 
$q \bar q, gg \to t \bar t h$, and $VV \to h$ refers to the vector
boson fusion processes at the LHC and the LC, 
$q \bar q \to q' \bar q' h$, $e^+e^- \to \bar \nu \nu h$,
respectively): 

\begin{itemize}

\item
$q \bar q, qq \to V \to V h$ ($V = W,Z$):\\
\BE
\frac{\si_{\rm hV}^{\SU}}{\si_{\rm hV}^{\SM}} 
 \approx \sin^2(\be - \aeff)
\label{qqVh}
\EE

\item
$gg \to h$:\\
\BE
\frac{\si^{\SU}(gg \to h)}{\si^{\SM}(gg \to h)}
 \approx \frac{\Ga^{\SU}(h \to gg)}{\Ga^{\SM}(h \to gg)}
\EE

\item
$q \bar q, gg \to t\bar t \to t \bar t h$:
\BE
\frac{\si^{\SU}(t \bar t \to t \bar t h)}{\si^{\SM}(t \bar t \to t \bar t h)}
 \approx \frac{\cos^2\aeff}{\sin^2\be}
\EE

\item
$VV \to h$, ($V = W, Z$):\\
\BE
\frac{\si_{\rm VVh}^{\SU}}{\si_{\rm VVh}^{\SM}} 
 \approx \sin^2(\be - \aeff)
\EE

\item
$e^+e^- \to Z \to Z h$:\\
\BE
\frac{\si_{\rm hZ}^{\SU}}{\si_{\rm hZ}^{\SM}}
 \approx \sin^2(\be - \aeff)
\EE

\item 
$\gagah$:\\
\BE
\frac{\si^{\SU}(\gagah)}{\si^{\SM}(\gagah)} 
 \approx \frac{\Ga^{\SU}(\hgaga)}{\Ga^{\SM}(\hgaga)}
\label{mumuh}
\EE

\item
$\mumuh$:\\
\BE
\frac{\si^{\SU}(\mumuh)}{\si^{\SM}(\mumuh)}
 \approx \frac{\Ga^{\SU}(\hmumu)}{\Ga^{\SM}(\hmumu)}
\EE

\item 
Higgs boson decays:\\
\BEA
&& \frac{\br^{\SU}(\hbb)}{\br^{\SM}(\hbb)}, \;
   \frac{\br^{\SU}(\hcc)}{\br^{\SM}(\hcc)}, \;
   \frac{\br^{\SU}(\htautau)}{\br^{\SM}(\htautau)}, \non \\
&& \frac{\br^{\SU}(\hgaga)}{\br^{\SM}(\hgaga)}, \;
   \frac{\br^{\SU}(\hWW)}{\br^{\SM}(\hWW)}, \;  
   \frac{\br^{\SU}(\hgg)}{\br^{\SM}(\hgg)}~.
\EEA

\end{itemize}

For some of the cross sections in \refeqs{qqVh} -- (\ref{mumuh}) more
complete results for the MSSM exist in the literature than those used
in our analysis, see e.g.\ \citeres{eehZhA,eennH,wiener,gghSUSY}. For the
qualitative analysis below, however, the approximations used here
should be sufficient, see e.g.\ the comparison in \cite{eehZhA}. 

For the numerical analysis the above production and decay modes
have been combined to the most relevant channels for each collider,
i.e.\ we have calculated the product
\BE
\frac{\si^{\SU}({\rm Higgs~prod.})}{\si^{\SM}({\rm Higgs~prod.})}
\times
\frac{\br^{\SU}({\rm Higgs~decay})}{\br^{\SM}({\rm Higgs~decay})} .
\label{master}
\EE
We take into account all possible Higgs boson decay channels including 
the full set of SUSY final states (in case the decay is kinematically
allowed). This includes the invisible decay into the lightest neutralino, 
$h \to \tilde{\chi}^0_1 \tilde{\chi}^0_1$. However, we have not found any
region of parameters in the three soft SUSY-breaking scenarios in which 
this decay channel becomes sizable. 
In the following, we list the relevant Higgs boson production and decay
channels for the various colliders:

\begin{itemize}

\item
Tevatron:
\BEA
q \bar q &\to& V^* \;\to\; V h \;\to\; V b \bar b\, , ~(V = W,Z)
\label{lhc:qqVh}
\EEA

\item
LHC:
\BEA
gg &\to& h \;\to\; \ga\ga \\
q \bar q, gg \;\to\; t \bar t &\to& t \bar t h \;\to\; t \bar t b \bar b \\
\label{lhc:VVh}
VV &\to& h \;\to\; \tau^+\tau^-, W W^*
\EEA

\item
LC:
\BEA
e^+e^- &\to& Z^* \;\to\; Z h \;\to\; Z b \bar b, Z c \bar c, 
                           Z \tau^+\tau^-, Z W W^*, Z gg \\
WW &\to& h \;\to\; b \bar b, c \bar c, \tau^+\tau^-, WW^*, gg
\EEA
Due to our approximations for the production processes the relative
results of\\ $\si^{\SU}/\si^{\SM} \times \br^{\SU}/\br^{\SM}$ for the
same final states in the two chains numerically agree.

\item
\gaC:
\BEA
\ga \ga &\to& h \;\to\; b \bar b, WW^*, \ga \ga
\EEA

\item
\muC:
\BEA
\mu^+ \mu^- &\to& h \;\to\; b \bar b, \tau^+\tau^-, WW^*
\EEA

\end{itemize}

%%%%%%%%%%%%%%%%%%%%%%%%%%%%%%%%%%%%%%%%%%%%%%%%%%%%%%%%%%%%%%
%%%%%%%%%%%%%%%%%%%%%%%%%%%%%%%%%%%%%%%%%%%%%%%%%%%%%%%%%%%%%%

\subsection{Phenomenological constraints}
\label{subsec:constraints}

While our main focus in this paper is on the physics in the Higgs
sector, we also take into account some further (relatively mild)
constraints when determining the allowed parameter space. These
constraints are briefly summarized here. (A more detailed discussion can be
found in \citere{asbs}.)

\begin{itemize}

\item
{\bf LEP Higgs bounds}:

The results from the Higgs search at LEP have excluded a considerable 
part of the MSSM parameter space~\cite{mssmhiggs}. 
The results of the search for the MSSM Higgs bosons are usually
interpreted in three different benchmark scenarios~\cite{benchmark}.
The 95\% C.L.\ exclusion limit for the SM Higgs boson of 
$\MHSM > 114.4 \gev$~\cite{LEPHiggs} 
applies also for the lightest $\cp$-even Higgs boson of the MSSM
except for the parameter region with small $\MA$ and large $\tb$. In the
unconstrained MSSM this bound is reduced to 
$\mh > 91.0 \gev$~\cite{mssmhiggs} for 
$\MA \lsim 150 \gev$ and $\tb \gsim 8$ as a consequence of a reduced
coupling of the Higgs to the $Z$~boson. For the $\cp$-odd Higgs boson a
lower bound of $\MA > 91.9 \gev$ has been obtained~\cite{mssmhiggs}.
In order to correctly interpolate between the parameter regions where
the SM lower bound%
\footnote{
Instead of the actual experimental lower bound, 
$\MHSM \gsim 114.4 \gev$~\cite{LEPHiggs},  
we use the value of $113 \gev$ in order to allow for some uncertainty
in the theoretical evaluation of $\mh$ from unknown higher-order
corrections, which is currently estimated
to be $\sim 3 \gev$~\cite{mhiggsAEC}.
}%
~of $\MHSM \gsim 113 \gev$ and the bound $\mh \gsim 91 \gev$
apply, we use the result for the Higgs-mass exclusion given with respect
to the reduced $ZZh$ coupling squared 
(i.e.\ $\sin^2(\be-\aeff)$)~\cite{Barate:2000zr}. We have compared the
excluded region with the theoretical prediction 
obtained at the \twol\ level for $\mh$ and $\sqbaeff$ for each parameter
set (using $\mt = 175 \gev$).

\item
{\bf Precision observables}:

The electroweak precision observables are affected by the whole spectrum
of SUSY particles. The main SUSY contributions enter via the
$\rho$-parameter~\cite{rhoparameter}. 
In our analysis we take into account 
the corrections arising from $\Stop/\Sbot$ loops up to two-loop
order~\cite{delrhosusy2loop}.
A value of $\De\rho$ outside the experimentally preferred region of 
$\dr^{\SU} \lsim 3\times 10^{-3}$~\cite{pdg} indicates experimentally
disfavored $\Stop$  and $\Sbot$ masses.
The evaluation of $\dr^{\SU}$ is implemented in \fh.

\item
{\bf Experimental bounds on SUSY particle masses}

In order to restrict the allowed parameter space in the three
soft SUSY-breaking scenarios we employed the current experimental 
constraints on their low-energy mass 
spectrum~\cite{pdg}. 
The precise values of the bounds that we have applied can be found in
\citere{asbs}. 

\item
{\bf Other restrictions}

We just briefly list here the further restrictions that we have taken
into account. For a detailed discussion see \citere{asbs}.

\begin{itemize}

\item 
The top quark mass is fixed to $\mt = 175 \gev$.

\item 
The GUT or high-energy scale parameters are taken to be real, no SUSY
$\cp$-violating phases are assumed. 

\item
In all models under consideration the $R$-parity symmetry~\cite{herbi}
is taken to be conserved.

\item
Parameter sets that do not fulfil the condition%
\footnote{We use here the one-loop minimization conditions. 
Analytical two-loop expressions of \order{\alpha_t\alpha_s+\alpha_t^2}
for the minimization conditions have been recently
given in \citere{ds}. 
The full two-loop corrections can be derived
numerically from the work of \citere{mhiggsEP2L}.
}%
of radiative electroweak symmetry breaking (REWSB) are discarded (already
at the level of model generation).

\item
Parameter sets that do not fulfil the constraints that there should
be no charge or color breaking minima are
discarded (already at the level of model generation).

\item
Contrary to \citere{asbs} we did not apply a ``naturalness bound'' on
the sfermion and gluino mass, but just restricted the scanned
parameter space as indicated in \refse{subsec:susybreak}.
This is in fact not an important restriction for the $\si \times \br$
calculation, since very heavy SUSY 
particles tend to decouple from the observables we are 
considering here, i.e.\ the quantity in \refeq{master} approaches~1.

\item
We demand that the lightest SUSY particle (LSP) is uncolored and uncharged.
In the mGMSB scenario the LSP is always the gravitino, so this 
condition is automatically fulfilled.
Within the mSUGRA and mAMSB scenario, the LSP
is required to be the lightest neutralino. 
Parameter sets that result in a different LSP are excluded.

\item
We do not apply any further cosmological constraints, i.e.\ we do not
demand a relic density in the region favored by dark matter
constraints~\cite{cdm}. 

\item
Although 
we do not apply constraints from $\br(b \to s \ga)$~\cite{bsgammaexp}
or $g_\mu - 2$ of the muon~\cite{gminus2}, we restrict ourselves to the
case where the Higgsino mixing parameter is positive, $\mu>0$. 
This choice is favored by the current data.
The results with negative $\mu$ can differ significantly from the case
we consider here and would require an additional analysis.

\end{itemize}

\end{itemize}

%%%%%%%%%%%%%%%%%%%%%%%%%%%%%%%%%%%%%%%%%%%%%%%%%%%%%%%%%%%%%%
%%%%%%%%%%%%%%%%%%%%%%%%%%%%%%%%%%%%%%%%%%%%%%%%%%%%%%%%%%%%%%

\subsection{Bounds on $\mh$ and $\tb$}
\label{subsec:mhtbbounds}

Scanning over the parameter space of mSUGRA, mGMSB and mAMSB as
described in \refse{subsec:susybreak} and applying the constraints as
described in \refse{subsec:constraints} results in upper maximal values
of $\mh$ and lower bounds on $\tb$ (for the general MSSM case, see
\citeres{mssmhiggs,tbexcl}). This analysis has been performed 
in \citere{asbs}. However, due to the progress in the $\mh$
evaluation, see \citere{mhiggsAEC} for a review, these bounds have
changed as compared to our earlier analysis. \refta{tab:mhtbbounds}
gives an update of the obtained $\mh$ and $\tb$ bounds.

%%%%%%%%%%%%%%%%%%%%% T A B L E %%%%%%%%%%%%%%%%%%%%%%%%%%%%%%%%%%%%%%%%%
\begin{table}[htb]
\begin{center}
\renewcommand{\arraystretch}{1.5}
\begin{tabular}{|l||c|c|c|} 
\cline{2-4} \multicolumn{1}{l||}{}
& mSUGRA & mGMSB & mAMSB \\ \hline \hline
$\mhmax$ [GeV] & 126.6  & 123.2 & 124.5 \\ \hline
$\tbmin$       & 2.9    & 3.2   & 3.8 \\ \hline \hline
\end{tabular}
\renewcommand{\arraystretch}{1.0}
\end{center}
\vspace{-0.5cm}
\caption{
Upper bound on $\mh$, $\mh < \mhmax$, and lower bound on $\tb$, 
$\tb > \tbmin$, in the three soft SUSY-breaking scenarios (for $\mt = 175
\gev$ and taking into account the phenomenological constraints of
\refse{subsec:constraints}; no theoretical uncertainties from unknown
higher-order corrections are included in $\mhmax$).
}
\label{tab:mhtbbounds}
\end{table}
%%%%%%%%%%%%%%%%%%%%% T A B L E %%%%%%%%%%%%%%%%%%%%%%%%%%%%%%%%%%%%%%%%%

%%%%%%%%%%%%%%%%%%%%%%%%%%%%%%%%%%%%%%%%%%%%%%%%%%%%%%%%%%%%%%
%%%%%%%%%%%%%%%%%%%%%%%%%%%%%%%%%%%%%%%%%%%%%%%%%%%%%%%%%%%%%%

\section{Observability of the lightest MSSM Higgs boson}
\label{sec:higgsobs}

In this section the observability of the lightest MSSM Higgs boson at
the different colliders is analyzed. Especially at the hadron
colliders a reduced $\si \times \br$ for certain channels 
compared to the SM value could
make it more difficult to establish a Higgs signal over the background.

Before we discuss the observability of the lightest Higgs boson at the
different colliders, we briefly summarize the main features of the
Higgs boson 
couplings in dependence on the relevant SUSY parameters in the three
soft SUSY-breaking scenarios. For a more extensive discussion see
e.g.~\citere{asbs}. The most important parameters are $\MA$ and $\tb$, 
since they enter the Higgs boson sector already at the tree-level. 
Deviations in the MSSM Higgs boson production and decay as compared
to the SM prediction arise in particular from modifications in the
$h\bar b b$ Yukawa coupling. 
The bottom and $\tau$ Yukawa coupling, being $\sim \Saeff/\Cb$, can be
strongly  
enhanced for small $\MA$ and large $\tb$ values. This can lead
to a strong enhancement of the partial widths $\Ga(\hbb)$, 
$\Ga(\htautau)$ and therefore also of the total Higgs boson
width. This gives rise to a corresponding 
suppression of the branching ratios of the other decay channels. 
In the case of mSUGRA, however, also a strong suppression of the
$h \bar b b$ and $h \tau^+\tau-$ Yukawa couplings could happen 
for small $\MA$, large $\tb$ and large $\mu$ due to radiative
corrections leading to a small value of $\Saeff$.

\begin{itemize}

\item Tevatron:\\
In all three soft SUSY-breaking scenarios the Tevatron search channels
are not significantly suppressed compared to the SM rates.
The channel $V \to V h \to V b \bar b$
is suppressed by not more than 10\% as compared to the SM value. 
Therefore the prospects at the Tevatron 
for the discovery of the lightest MSSM Higgs boson are as good as for
the SM Higgs boson. A similar observation has already been made in
\citere{ehow} for the mSUGRA scenario, where also other
phenomenological constraints have been taken into account.

\item LHC:\\
We start our discussion with the channel $gg \to h \to \ga\ga$ that is
of particular importance for a light SM-like 
Higgs boson with $\mh \lsim 130 \gev$~\cite{atlastdr}. As has been shown in
\citere{gghggsupp,belanger,benchmark2}, for certain values of the SUSY 
parameters this
production channel can be heavily suppressed for a wide region of the
$\MA-\tb$-parameter space of the unconstrained MSSM (see also
\citere{richterwas} for an analysis in the mSUGRA scenario). 
Since the event rate for this channel is relatively low ($\hgaga$
being a rare decay), a suppression of 50\% or more would certainly
pose a challenge to the experiment.

%%%%%%%%%%%%%%%%%%%%%%%%%%
\begin{figure}[ht!]
\vspace{-1em}
\begin{center}
\epsfig{figure=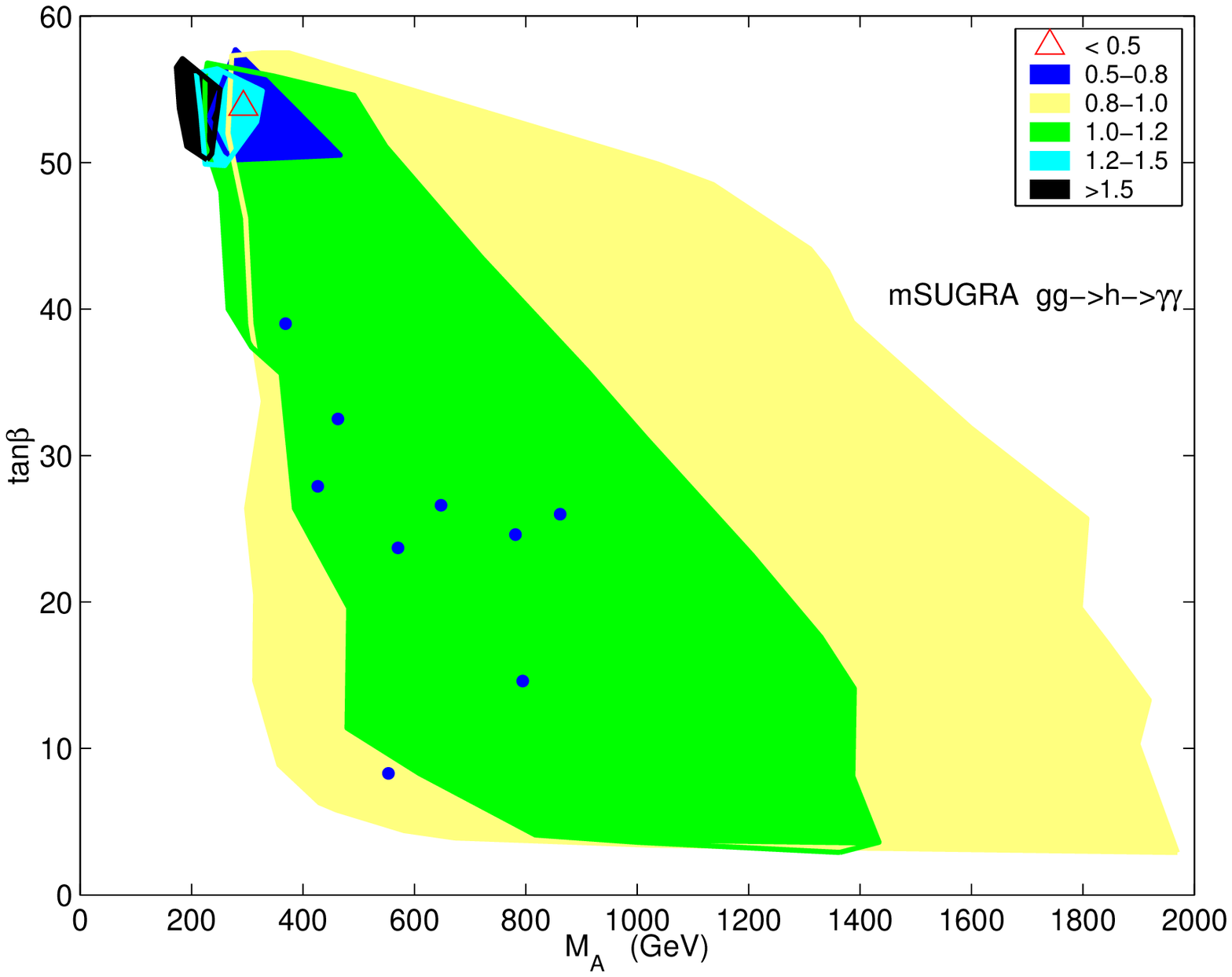,width=12cm,height=6.5cm}
\vspace{1em}
\epsfig{figure=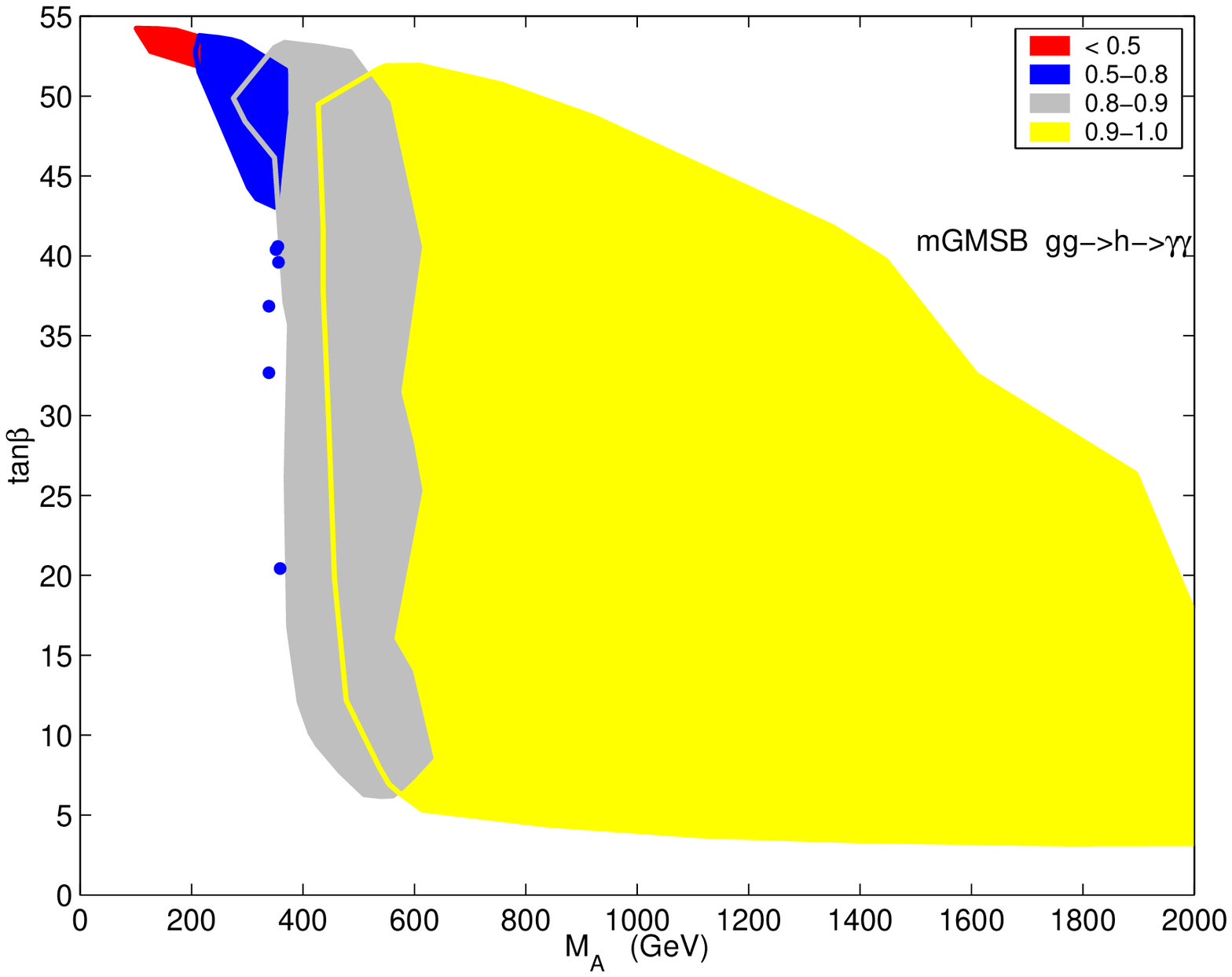,width=12cm,height=6.5cm}
\vspace{1em}
\epsfig{figure=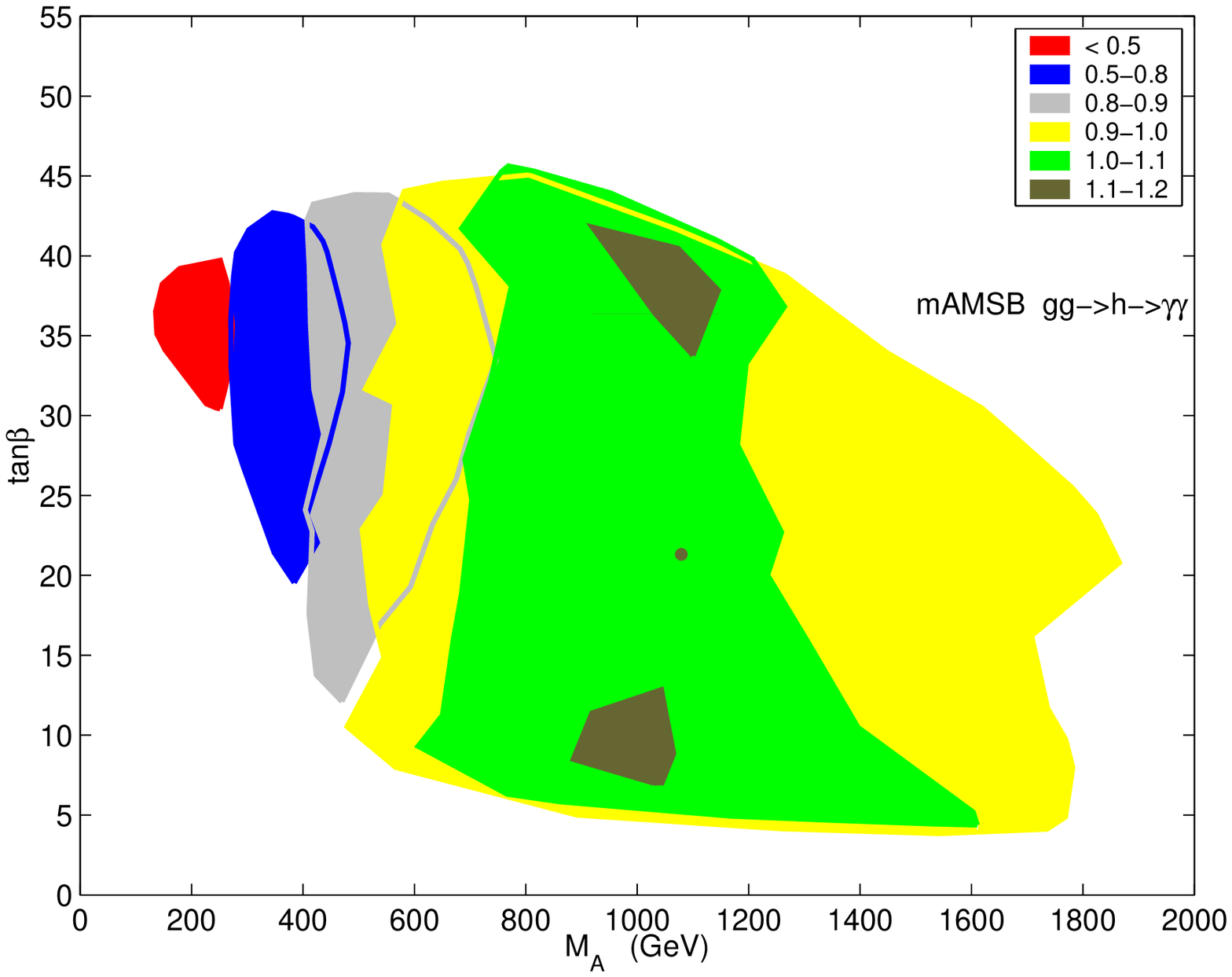,width=12cm,height=6.5cm}
\caption{
The rate for the LHC channel $gg \to h \to \ga\ga$, normalized to the SM 
case with the same Higgs boson mass, 
$(\si \times \br)^{\SU}/(\si \times \br)^{\SM}$,
is shown in the $\MA$--$\tb$ plane
for the mSUGRA, mGMSB and mAMSB scenarios (from top to bottom). 
The unconnected dots appearing in this and the following figures
reflect the fact that only a few points with very low density appear
in this area. However, increasing the density of the scatter data would
cover the whole area in which the dots are located.
}
\label{ggggASBS}
\end{center}
\vspace{-3em}
\end{figure}
%%%%%%%%%%%%%%%%%%%%%%%%%

The situation within the three soft SUSY-breaking scenarios can be
read off from \reffi{ggggASBS}~%
\footnote{
For a similar analysis for the charged Higgs bosons, see
\citere{sola}.
}%
. Within the mSUGRA scenario
a suppression of up to 10--20\% is found for $\MA \lsim 700 \gev$ for
all $\tb$ values.
For very large $\tb$, $\tb \gsim 50$, a reduction even
larger than 20\% can be found.  
This suppression is due to the fact that the $\br(\hbb)$ is strongly
enhanced in this part of the parameter space, as explained in the beginning
of this section. 
This is in agreement
with the result obtained in
\citere{ehow}, where no substantial reduction has been found, after
other phenomenological constraints like $\br(b \to s \ga)$, 
$g_\mu - 2$ and especially the CDM restrictions had been applied.
Concerning the parameter space of the GUT scale 
parameters $m_0$ and $m_{1/2}$ (which
is not shown here explicitly), the reduction is found for 
all $m_0$ and $m_{1/2} \lsim 350 \gev$.
It should be noticed that an enhancement of 50\% or more is possible
in the very  small $\MA$ region due to a possible strong suppression
of $\hbb$ in accordance with the analysis in \citere{asbs}.

The suppression can be stronger in the other two scenarios. 
Nearly all model points with $\MA \lsim 400 \gev$ 
for mGMSB and mAMSB, which correspond to $\tb$ values of $\tb \gsim 20$, 
show a suppression of 20-50\% (or even more).
For $\MA \lsim 200 \gev (300 \gev)$ and 
$\tb \gsim 50$ ($30 \lsim \tb \lsim 40$) 
the reduction is even larger than 50\% in mGMSB (mAMSB). 
For larger $\MA$ values, $\MA \gsim 600 \gev$, the 
SM value of $\si \times \br$ is approached. 
Concerning the high-energy parameters, the largest reduction in the
mGMSB scenario is found all over the parameter space
for $M_{\rm mess}$ and $N_{\rm mess}$ (with 
$\tb \gsim 50$ and $\La \gsim 30 \tev$), 
whereas a reduction by 10-20\% is mostly found for the lowest 
$\La$ values for all $M_{\rm mess}$ values. 
Within mAMSB the largest reduction is found for
$m_0 \lsim 700 \gev$ and $m_{\rm aux} \lsim 4 \times 10^4 \gev$. 
Small values of the high-energy parameters correspond to relatively
small values of the low-energy SUSY parameters, which are required for a
sizable suppression of the $gg \to h \to \ga\ga$ 
channel~\cite{gghggsupp,belanger,benchmark2}.
The fact that the suppression of this channel can be much more
pronounced in mGMSB and mAMSB as compared to mSUGRA originates to a
large extent from the different behavior of the $\br(\hbb)$ channel
(see also below). This dominant decay channel can be much more
strongly enhanced in mGMSB and mABSB~\cite{asbs} and thus suppress the
decay of the lightest Higgs boson to photons.

It is interesting to note that only in the mAMSB scenario an
enhancement of the $gg \to h \to \ga\ga$ can be found in the
intermediate $\MA$ region, which is absent in the other scenarios
(within mSUGRA the region of enhancement labeled with ``1.0 -- 1.2'' in
general has values very close to 1.0).
The reason for this enhancement is the following combination of effects:
for the intermediate $\MA$ region, the partial decay width
$\Ga(\hgaga)$ can be slightly enhanced due to loop corrections
in mSUGRA and mGMSB (less then
10\%), while it is more enhanced in mAMSB (up to 20\%). 
The total decay width is dominated by $\Ga(\hbb)$, which can be
enhanced 
up to $10-20\%$ in mGMSB, but 10\% at most within the
mAMSB scenario. This results in a suppression (or at most only a very
mild increase) of $\br(\hgaga)$ within mSUGRA and
mGMSB as compared to the SM, but in an enhancement for mAMSB in the
intermediate $\MA$ region.

\smallskip
We now consider the associated production channel at the LHC,
$q \bar q, gg \to t\bar t \to t \bar t h \to t \bar t b \bar b$.%
\footnote{
Also some other channels for a SM Higgs boson have been studied,
e.g.\ $t \bar t \to t \bar t h \to t \bar t WW^*$~\cite{tthWW}. They
could be easily included in our analysis as outlined in
\refse{sec:higgs}.  Since including these channels does not change our
qualitative results, we do not discuss them here explicitly.
}%
~In all three scenarios
the rate is not suppressed compared to the SM rate by more than 10\% 
or even an enhancement by up to
10-20\% occurs. Therefore this channel, which is most relevant in the region 
$100 \gev \lsim$ $\mh$ $\lsim 120 \gev$~\cite{atlastdr}, does not suffer
from suppression due to SUSY corrections in the three soft SUSY-breaking
scenarios.
More severe suppressions are possible for the 
$q \bar q, gg \to b \bar b h$ channel. This channel can be observed for very
small $\MA$ and large $\tb$, but plays only a minor role
concerning the observability of the lightest MSSM Higgs boson at the
LHC~\cite{schumitalkPrag}. 

\smallskip
More recently also the Higgs boson production via $W$~boson fusion,
$W^+W^- \to h$, with a subsequent decay to $\tau^+\tau^-$ pairs,
$W$~bosons or photons has been discussed,
see \citeres{zeppi,plehnix3} for a SM analyses and
\citeres{WWhtautau,plehnix2} for the mode
$\htautau$ in the MSSM case.
These channels can be relevant for the whole allowed 
$\mh$ range in the MSSM. The mSUGRA scenario offers very good prospects for
the decay into $\tau^+\tau^-$, which is 
characterized by the coupling $\sim \Saeff/\Cb$. Over almost 
the whole parameter space the rate for $\si \times \br$ differs by less
than 10\% from the SM rate. 
Only for very small $\MA$ and very high $\tb$ a suppression is
possible. This corresponds to the parameter space where the heavy
Higgs boson can have SM like couplings, see Fig.~1 in
\citere{asbs}. The other channels ($h \to WW^*, \gamma\gamma$) show the
following pattern: due to the increased decay rates to fermions, at relatively
small $\MA$, $250 \gev \lsim \MA \lsim 500 \gev$, a suppression of the
decays into $W$~bosons or photons is possible, while the 
value of the SM rate is approached for larger $\MA$.

%%%%%%%%%%%%%%%%%%%%%%%%%%
\begin{figure}[ht!]
\begin{center}
\epsfig{figure=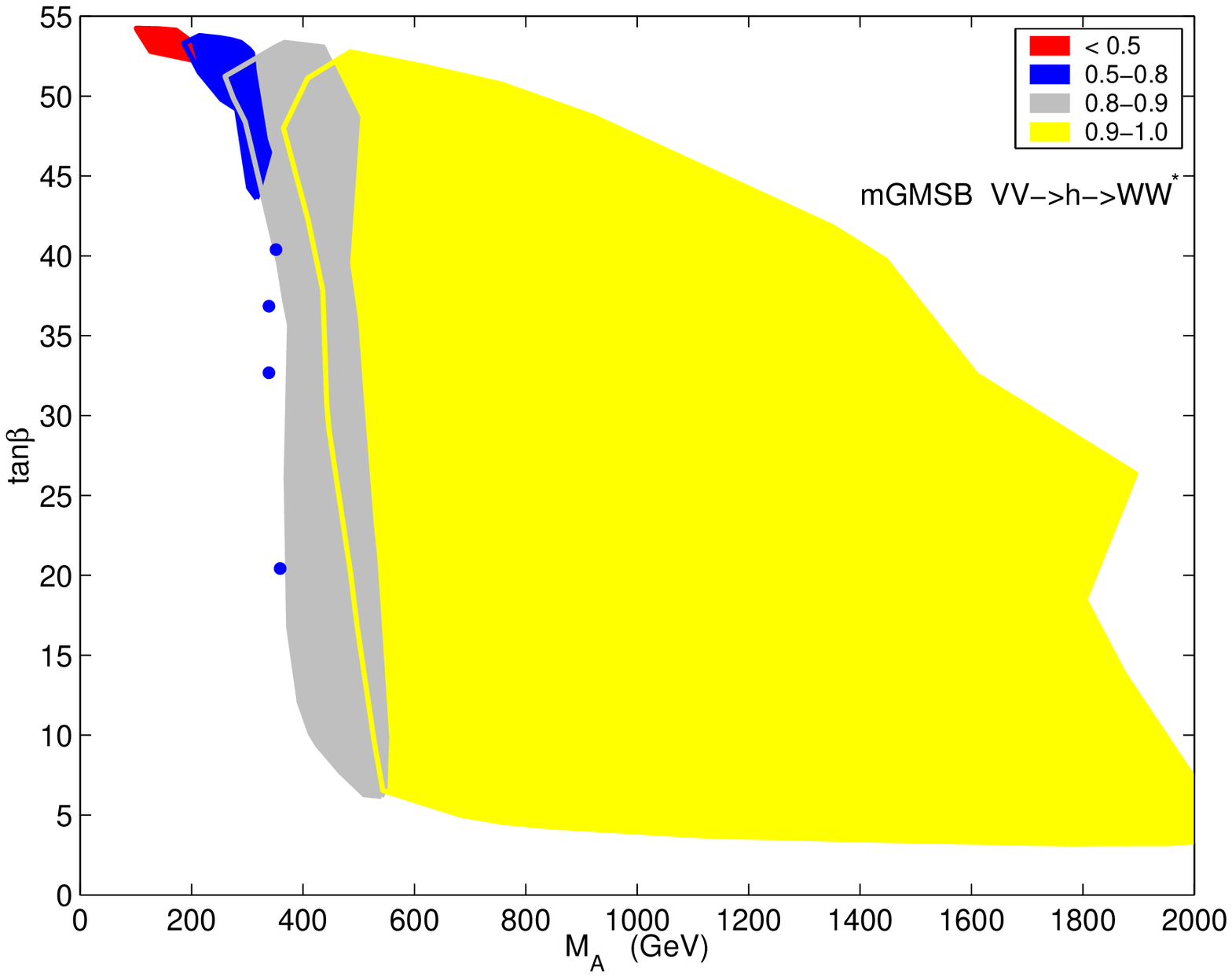,width=12cm,height=6.5cm}
\vspace{1em}
\epsfig{figure=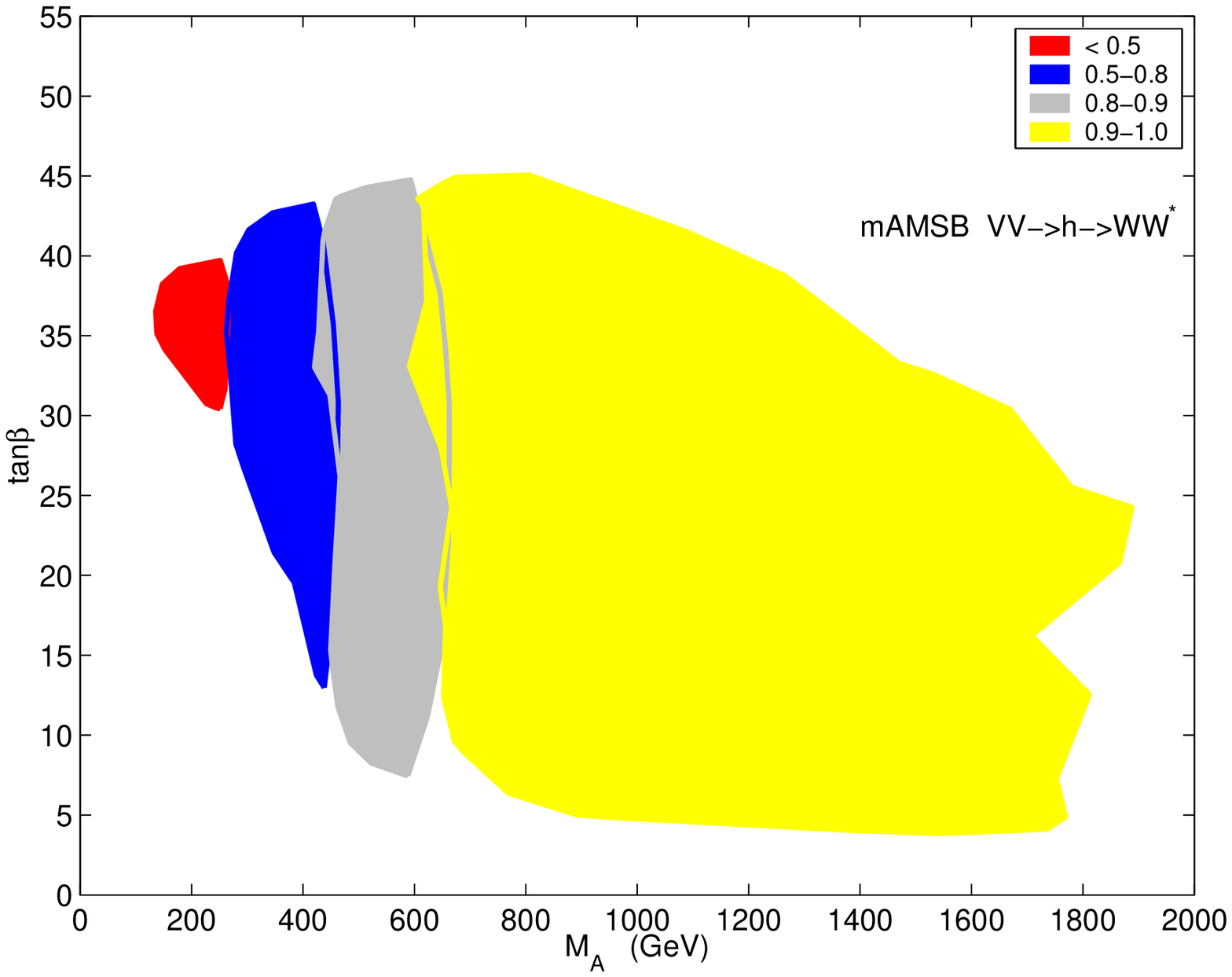,width=12cm,height=6.5cm}
\caption{
The rate for the LHC channel $WW \to h \to WW^*$, normalized to the SM 
case with the same Higgs boson mass, is shown in the $\MA$--$\tb$ plane
for the mGMSB (upper plot) and the mAMSB scenario (lower plot). 
}
\label{VVWW_GA}
\end{center}
\vspace{-1em}
\end{figure}
%%%%%%%%%%%%%%%%%%%%%%%%%

The situation is similar in the mGMSB and mAMSB scenarios, besides that
the suppression of the $\tau^+\tau^-$ channel for very small $\MA$ and
very large $\tb$ is not found. For the smallest possible values of $\MA$ and
the largest possible values of $\tb$ a strong enhancement of these
channels can be observed. This agrees with the results of
\citere{asbs}, where only in the mSUGRA scenario but not in the mGMSB
and mAMSB scenario a region of the parameter space has been found
where the heavy Higgs boson is SM like. Thus the $\htautau$
channel is enhanced everywhere in mGMSB and mAMSB. Correspondingly
the two other channels, $\hWW$ and $h \to \gamma\gamma$ are suppressed
everywhere. As an example, in \reffi{VVWW_GA} we show the channel 
$WW \to h \to WW^*$ in the mGMSB and the mAMSB scenario. For not too 
large $\MA$, $\MA \lsim 550 \gev (700 \gev)$, in mGMSB (mAMSB) a reduction 
larger than 10\% can be observed. For larger $\MA$ the results in the
mGMSB and mAMSB scenarios approach the one in the SM.

\item LC:\\
Due to its clean experimental environment, Higgs boson production should 
be easily observable at the LC, i.e.\ in various channels even a significant 
suppression compared to the SM rate would not be 
harmful~\cite{teslatdr,orangebook,acfarep}. Therefore the
production channels $Z^* \to Z h$ and $WW \to h$ (which
yield the same numerical result in our analysis, see
\refse{subsec:mssmhiggs}) and all decay channels, $h \to b \bar b$, 
$c \bar c$, $\tau^+\tau^-$, $WW^*$, $gg$, are observable in the three
soft SUSY-breaking scenarios. This applies also in the region of small
$\MA$, $\MA \lsim 200 \gev$, where a suppression of more than 50\% can
occur in all three scenarios. We will therefore present a detailed
analysis of the LC production and decay channels only in the context of
precision measurements, see \refse{sec:higgsbrs}.

\bigskip
\item \gaC:\\
Also the \gaC , due to the Higgs boson production in the s-channel,
offers very good prospects for the Higgs boson
observation~\cite{gagahiggs,MAdet1}. Only the decay $\hgaga$ could become
problematic if it is strongly suppressed compared to the SM value
(e.g.\ due to an enhanced $hb\bar b$ coupling). 

%%%%%%%%%%%%%%%%%%%%%%%%%%
\begin{figure}[ht!]
\begin{center}
\epsfig{figure=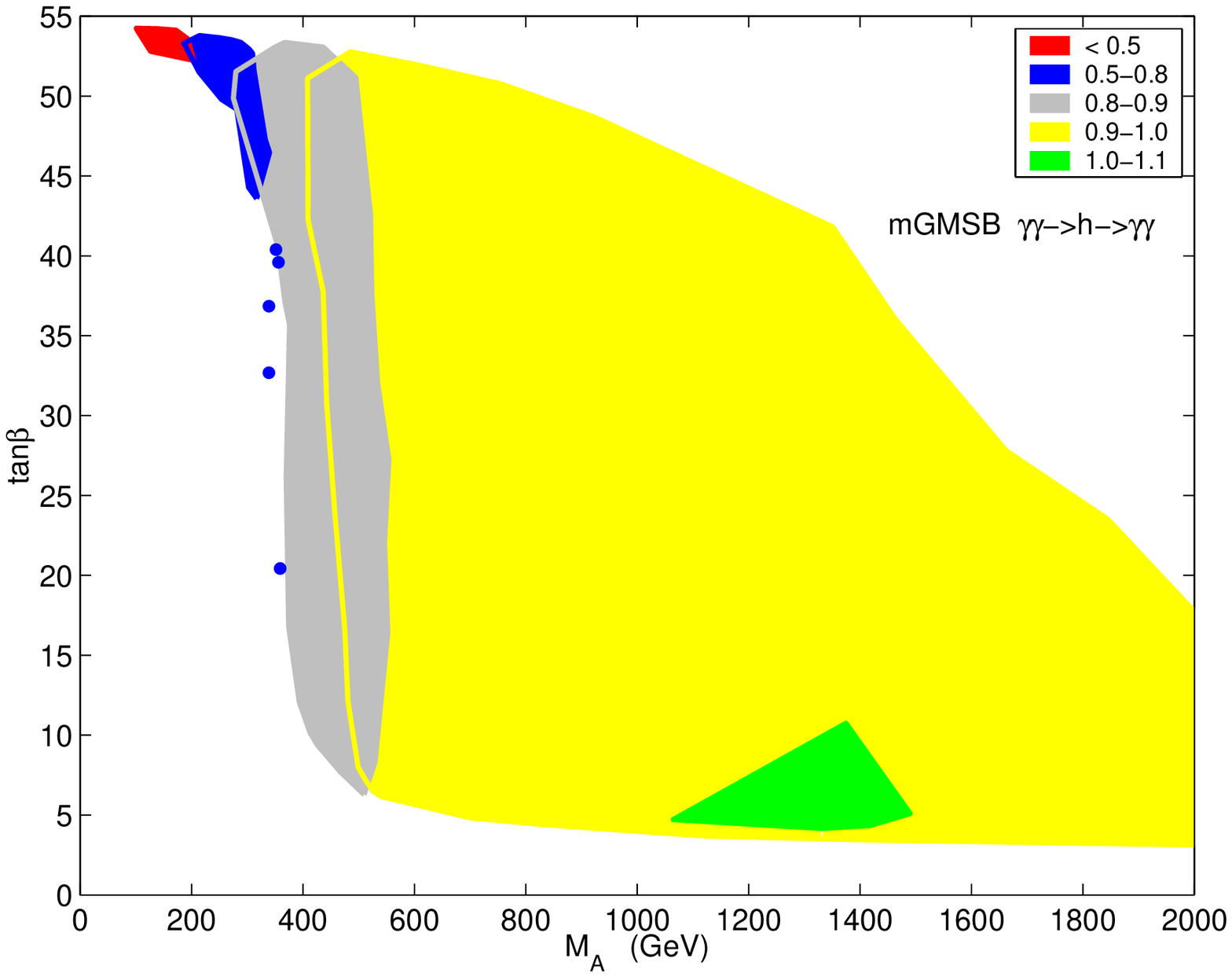,width=12cm,height=6.5cm}
\vspace{1em}
\epsfig{figure=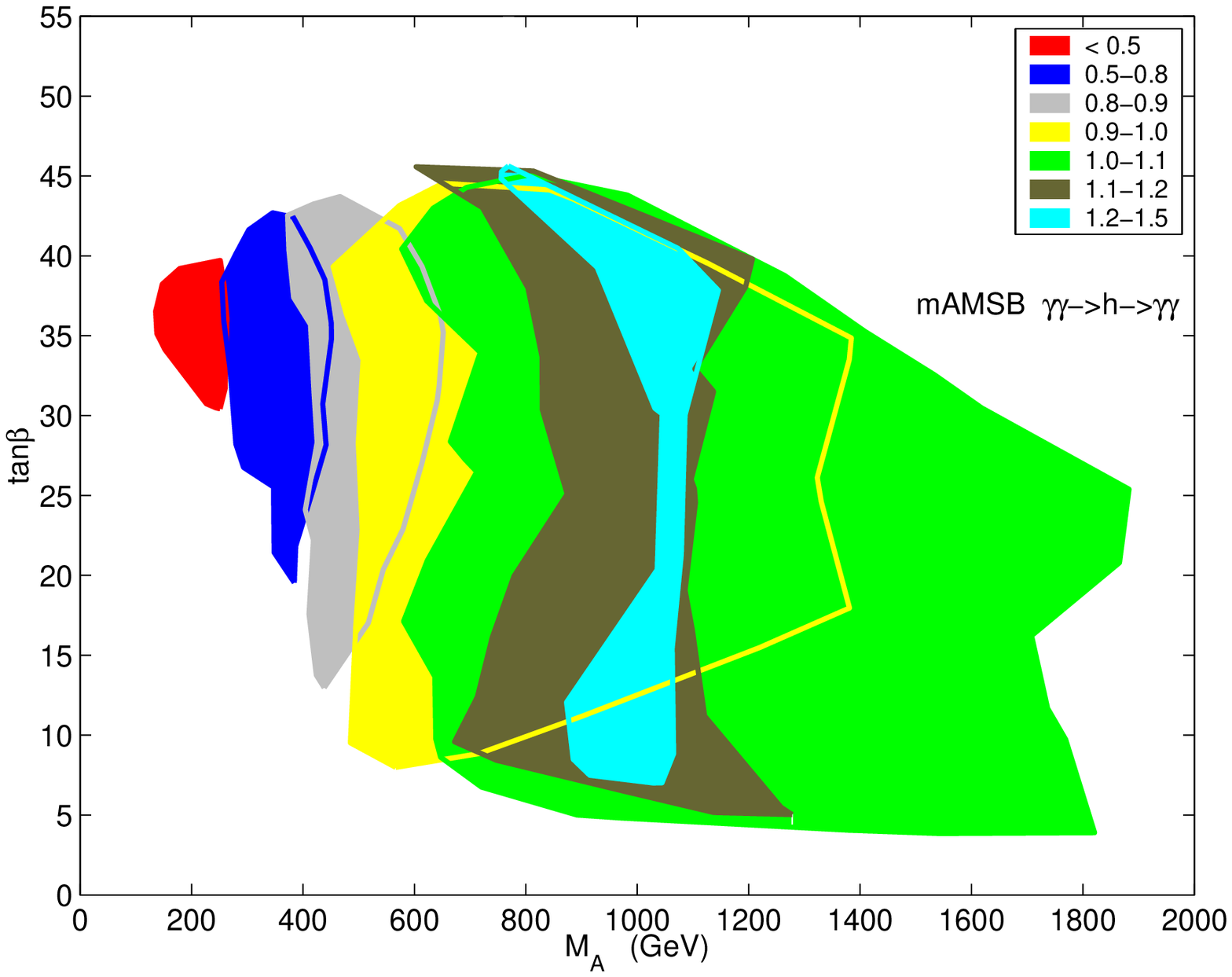,width=12cm,height=6.5cm}
\caption{
The rate for the \gaC\ channel $\ga\ga \to h \to \ga\ga$, normalized to the SM 
case with the same Higgs boson mass, is shown in the $\MA$--$\tb$ plane
for the mGMSB (upper plot) and the mAMSB scenario (lower plot). 
}
\label{gaga_GA}
\end{center}
\vspace{-1em}
\end{figure}
%%%%%%%%%%%%%%%%%%%%%%%%%

In the mSUGRA scenario the $\hgaga$ channel and the $\hWW$ channel are
very similar. They can be suppressed by more than
20\% only for very large $\tb$, $\tb \gsim 50$, or for $\tb \approx 10$
with the smallest allowed $\MA$ values. 
On the other hand, the $\hbb$ and $\htautau$ modes are unproblematic
within mSUGRA and show either only a very small suppression or even some
enhancement, with the only exception of a possible suppression at
very large $\tb$.  

In the mGMSB scenario the $\hbb$ and $\htautau$ channels are always 
enhanced. The $\hgaga$ and $\hWW$ channels show a suppression of 
more than 50\% for $\MA \lsim 100 \gev$, while for $\MA \lsim 600
\gev$ still a suppression of more than 10\% occurs,
see \reffi{gaga_GA} for the $\hgaga$ channel.
The situation is quite similar in the mAMSB scenario,
with an exception in the intermediate $\MA$ region, $600 - 1300 \gev$,
where there is an enhancement of the $\hgaga$ mode compared to the
one obtained in the mGMSB scenario, see \reffi{gaga_GA}.
The reason for this enhancement is similar to the case of the
$gg \to h \to \ga\ga$ channel at the LHC (see also the discussion of 
\reffi{ggggASBS}).
There the branching ratio for $\hgaga$ is enhanced in mAMSB as compared to
mGMSB (and mSUGRA). At the \gaC\ the effect is even more pronounced
since the enhanced $\ga\ga h$ vertex now also enters in the Higgs
production. Correspondingly, the $\ga\ga \to h \to \ga\ga$ channel can
be only slightly enhanced in mGMSB, but more strongly increased in the
mAMSB scenario. 
For the process $VV \to \hWW$ that is important at the LHC and the LC,
see e.g.\ \reffi{VVWW_GA}, the mGMSB and the mAMSB scenario are very
similar, since no $h\ga\ga$ vertex is involved.

\item \muC:\\
Finally, to complete our analysis, we also briefly look at the
\muC. This collider offers good prospects since the Higgs
boson can be produced in the s-channel without a loop suppression like
at the \gaC, however with the relatively small $\mu^+\mu^-h$ Yukawa
coupling~\cite{mumuhold,mumuhiggs}. 
The production of SUSY Higgs bosons at the \muC\ has been extensively
discussed in the literature, see e.g.\ \citeres{mumuhold,mumuhiggs}
and references therein, but the impact of the different SUSY-breaking
scenarios has not been investigated yet. 
In the unconstrained MSSM 
it is possible that the $\mu^+\mu^-h$ coupling, being $\sim \Saeff/\Cb$,
can become very small if $\aeff \to 0$ because of loop
corrections~\cite{mumuhiggs,mumuhsup}. In this parameter region, on the
other hand, $H$, $A$ production at the \muC\ happens with
an enhanced rate and offers good prospects for resolving~$H$ and~$A$ as
separate resonances~\cite{mumuhiggs,mumuhsup}.

The feature of a suppressed $\mu^+\mu^-h$ coupling
can also be realized in the mSUGRA scenario when the heavy
(and not the light) $\cp$-even Higgs boson is SM like. This is
possible for very high $\tb$ and small $\MA$, $\MA \lsim 300 \gev$ (note
that here no CDM constraints are taken into account, in contrast to the
analysis of \citere{mumuhiggs}). In
this parameter region a strong suppression is possible for 
all \muC\ channels of the light $\cp$-even Higgs boson (while $H$, $A$ 
production happens at enhanced rates).
In the rest of the parameter space the $\hbb$ and
$\htautau$ channel are strongly enhanced for $\MA \lsim 700 \gev$. For
very large $\MA$ an enhancement of up to 10\% occurs.
Correspondingly, the
$\hWW$ channel is not enhanced, but still within 10\% of the SM value.

Within mGMSB and mAMSB the suppression of the $\mu^+\mu^-h$ coupling
is not present. Because of the coupling factor $\Saeff/\Cb$ the $\hbb$
and $\htautau$ channels are strongly enhanced for small $\MA$, while the
SM value is approached for large values of $\MA$. 
The $\hWW$ channel, being enhanced with the
$\Saeff/\Cb$ factor only at the production vertex, is less enhanced,
but should be unproblematic for the whole parameter space.

\end{itemize}

%%%%%%%%%%%%%%%%%%%%%%%%%%
\begin{table}[ht!]
\hspace{-3cm}
\psfig{figure=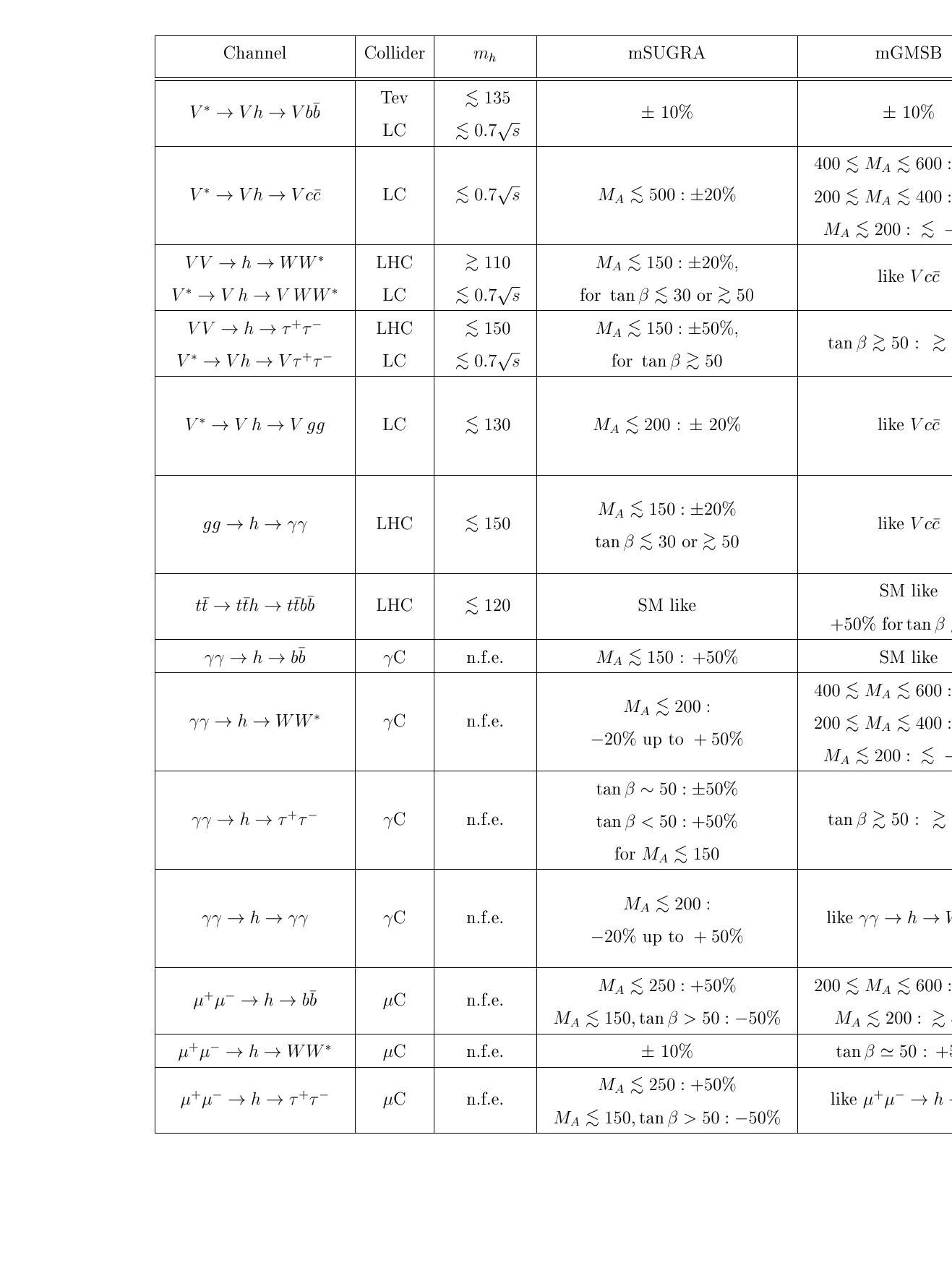,width=14cm}
\vspace{-2cm}
\caption{
Behavior of the production and decay modes of the lightest $\cp$-even Higgs
boson of the MSSM,
see \refeq{master},
for the most relevant channels at present and
future colliders in three
different SUSY breaking scenarios.
When there is suppression or enhancement  we indicate its
{\it maximum} magnitude together with the parameters where this happens.
If not stated explicitly, for the rest of the parameter
space the mode behaves roughly like in the SM. The Higgs mass range where
detection of a statistically significant signal is possible is also
shown~\cite{LHCreach,LCreach}. The phrase ``n.f.e.'' stands for ``not
fully explored'' and refers to channels where the
studies so far have been performed for some fixed $\mh$~values only.
}
\label{tab:summary}
\end{table}
%%%%%%%%%%%%%%%%%%%%%%%%%%%%%%%%%%%%%%%%%%%%

\smallskip
The results of this section are summarized in \refta{tab:summary}.
The modes $gg \to h \to \ga\ga$, $t\bar t \to t\bar t h$ and 
$WW \to h \to WW^*, \tau^+\tau^-, \ga\ga$ allow the detection of the
lightest MSSM 
Higgs boson in all three scenarios over the whole indicated
parameter space.
Possible exceptions occur for the $gg \to h \to \ga\ga$ channel for
the very small $\MA$ region in the mGMSB and mAMSB scenarios, 
where a strong suppression of more than 50\% could happen.
Because of the clean experimental environment at a LC, detection of the
light Higgs is ensured for all the three scenarios. 
For a \gaC, the region of very small $\MA$ values in the mGMSB
and mAMSB scenarios might be difficult for the $\hWW$ and $\hgaga$ mode
because of a strong suppression of more than 50\%. 
At the \muC, the region $\MA \lsim 150 \gev$ and $\tb \gsim 50$ in the mSUGRA
scenario exhibits a significant suppression of the production of the
lightest $\cp$-even MSSM Higgs boson. Besides the ``difficult'' regions
mentioned above, the main search modes at the \gaC\ and the
\muC\ are not affected by strong suppression for all three scenarios,
and the lightest MSSM Higgs boson in these scenarios will 
clearly be detectable at all possible future colliders.

%%%%%%%%%%%%%%%%%%%%%%%%%%%%%%%%%%%%%%%%%%%%%%%%%%%%%%%%%%%%%%
%%%%%%%%%%%%%%%%%%%%%%%%%%%%%%%%%%%%%%%%%%%%%%%%%%%%%%%%%%%%%%

\section{Precision analyses of the Higgs masses and branching ratios}
\label{sec:higgsbrs}

We now investigate the potential of Higgs branching ratio measurements at 
future colliders for testing the underlying SUSY model. We concentrate
our analysis on the LC and the \gaC, since the anticipated precisions of
at the LHC will in general be much worse, while on the other hand
branching ratio measurements of the light Higgs boson 
at the \muC\ are not expected to yield substantially better results than
at the LC.

Over most of the parameter space of the scenarios
discussed here one would of course also expect to observe direct
production of SUSY particles at the next generation of colliders.
However, we concentrate our analysis on information obtainable from the
Higgs sector without assuming further knowledge of the SUSY spectrum. 
In a realistic situation one would of course confront the model under
study with all available experimental information.

\refta{tab:BRunc} 
lists the anticipated accuracies in different channels at the
LC~\cite{teslatdr,talkbrient} and the \gaC~\cite{gagahiggs,MAdet1}.
The values given in \refta{tab:BRunc} correspond to a SM-like Higgs
boson with a mass compatible with the allowed mass range of the lightest
$\cp$-even Higgs boson in the three soft SUSY-breaking scenarios
according to the upper bounds given in \refta{tab:mhtbbounds}.
There is of course some variation in the accuracy with which the
branching ratios can be measured over the allowed range of $\mh$. For
simplicity, we assume a constant precision over the allowed mass range
for each channel with a value referring to the middle of the allowed
range.
In parameter regions where the MSSM rate differs drastically from the SM
rate the prospective precision will of course be different than in the
SM. While in extreme cases like this it will be easy to infer properties
of the SUSY model from Higgs sector measurements, we will focus in our
analysis below on moderate deviations between the MSSM and the SM, for
which the values given in \refta{tab:BRunc} can be applied in good
approximation.
We will indicate the deviation between the MSSM and the SM in terms of
the accuracies given in \refta{tab:BRunc}, i.e.\ a ``$\pm n\,\si$'' 
deviation means that the calculated MSSM value of $\si \times \br$
deviates from the corresponding SM value (with $\MHSM = \mh$) by 
$(\pm n \times {\rm precision})$.

%%%%%%%%%%%%%%%%%%%%% T A B L E %%%%%%%%%%%%%%%%%%%%%%%%%%%%%%%%%%%%%%%%%
\begin{table}[hb!]
\begin{center}
\renewcommand{\arraystretch}{1.5}
\begin{tabular}{|c|l|c|} \hline\hline
collider & decay mode & precision \\ \hline \hline
LC   & $\hbb$     &  1.5\% \\ \hline
LC   & $\htautau$ &  4.5\% \\ \hline
LC   & $\hcc$     &  6\%   \\ \hline
LC   & $\hgg$     &  4\%   \\ \hline
LC   & $\hWW$     &  3\%   \\ \hline \hline
\gaC & $\hbb$     &  2\%   \\ \hline
\gaC & $\hWW$     &  5\%   \\ \hline 
\gaC & $\hgaga$   &  11\%  \\ \hline \hline
\end{tabular}
\renewcommand{\arraystretch}{1.0}
\end{center}
\caption{Anticipated precisions for measurements of Higgs branching
ratios at the LC~\cite{teslatdr,talkbrient} and the 
\gaC~\cite{gagahiggs,MAdet1}. The values are given for a SM-like Higgs boson
with a mass compatible with the
allowed mass range of the lightest $\cp$-even Higgs boson in the three
soft SUSY-breaking scenarios, see text.
}
\label{tab:BRunc}
\end{table}
%%%%%%%%%%%%%%%%%%%%% T A B L E %%%%%%%%%%%%%%%%%%%%%%%%%%%%%%%%%%%%%%%%%

%%%%%%%%%%%%%%%%%%%%%%%%%%%%%%%%%%%%%%%%%%%%%%%%%%%%%%%%%%%%%%
%%%%%%%%%%%%%%%%%%%%%%%%%%%%%%%%%%%%%%%%%%%%%%%%%%%%%%%%%%%%%%

\subsection{Sensitivity to $\MA$ and $\tb$}
\label{subsec:MAsensitivity}

\smallskip
While within the MSSM the prospects for the detection of the lightest
$\cp$-even Higgs boson at the next generation of colliders are very
good, the situation is quite different for direct observation of the
$\cp$-odd $A$ boson.
At the LHC the detection of this particle can be very difficult over
sizable fractions of the MSSM parameter space (see e.g.\
\citere{atlastdr,cms}), while it may be outside the kinematical reach of the 
LC (see \citeres{eennH,eennHproc} for a recent account of this
subject). Thus, it is of 
interest to study the potential for 
obtaining indirect bounds on $\MA$ from precision measurements.
Exploiting the sensitivity to $\MA$ can be done in a similar fashion as 
nowadays for the SM Higgs, where indirect bounds are derived from
electroweak precision tests. Since in the decoupling limit, $\MA \gg
\MZ$, the Higgs sector of the MSSM becomes SM-like, deviations in the
production and decay of the lightest $\cp$-even Higgs boson of the MSSM
can in principle be translated into an upper bound on $\MA$. If direct
information on $\MA$ is available, the indirect sensitivity to $\MA$
allows a stringent test of the model.

\smallskip
Several analyses of the sensitivity to $\MA$ at the LC or the \gaC\
have been carried out in the literature~\cite{ehow,MAdet1,MAdet2,MAdet3,MAdet4}
(for an analysis focusing on the measurements with a GigaZ option of the
LC see \citere{gigaz}). While in many of these analyses particular
``benchmark'' values of the SUSY parameters have been chosen, we perform
a detailed scan over the parameter space of the three soft SUSY-breaking
scenarios. 
This is in contrast to previous studies on the Higgs branching ratios in the
literature~\cite{MAdet1,MAdet2,MAdet3}, where
all parameters except for the one under investigation have been kept
fixed. In this case the 
assumed deviation between the MSSM and the SM is solely attributed to
this single free parameter. This corresponds to a situation with 
a complete knowledge of all other SUSY parameters without any experimental or
theoretical uncertainty, which obviously leads to an unrealistic
enhancement of the sensitivity to the investigated parameter. Allowing
the other SUSY (and SM) parameters to vary within reasonable ranges would 
result in reduced sensitivities as compared to the ones reported in
these studies.

\smallskip
Since assumptions about which part of the SUSY spectrum might be
accessible at the next generation of colliders are necessarily very
speculative, we do not assume any further information beyond the Higgs
sector at all and perform a full scan over the parameter space of the
three soft SUSY-breaking scenarios. The resulting sensitivity to $\MA$
(which effectively covers also possible theoretical uncertainties%
\footnote{Note that the presently largest theoretical uncertainty in the
MSSM Higgs sector, which arises from the experimental error of the
top-quark mass, will be drastically reduced by the precise measurement
of $\mt$ at the LC.}
)
can thus be interpreted as a ``worst case'' scenario within mSUGRA, mGMSB
and mAMSB, which could be improved by incorporating further information
from other sectors of the model.

%%%%%%%%%%%%%%%%%%%%%%%%%%
\begin{figure}[ht!]
\vspace{-1em}
\begin{center}
\epsfig{figure=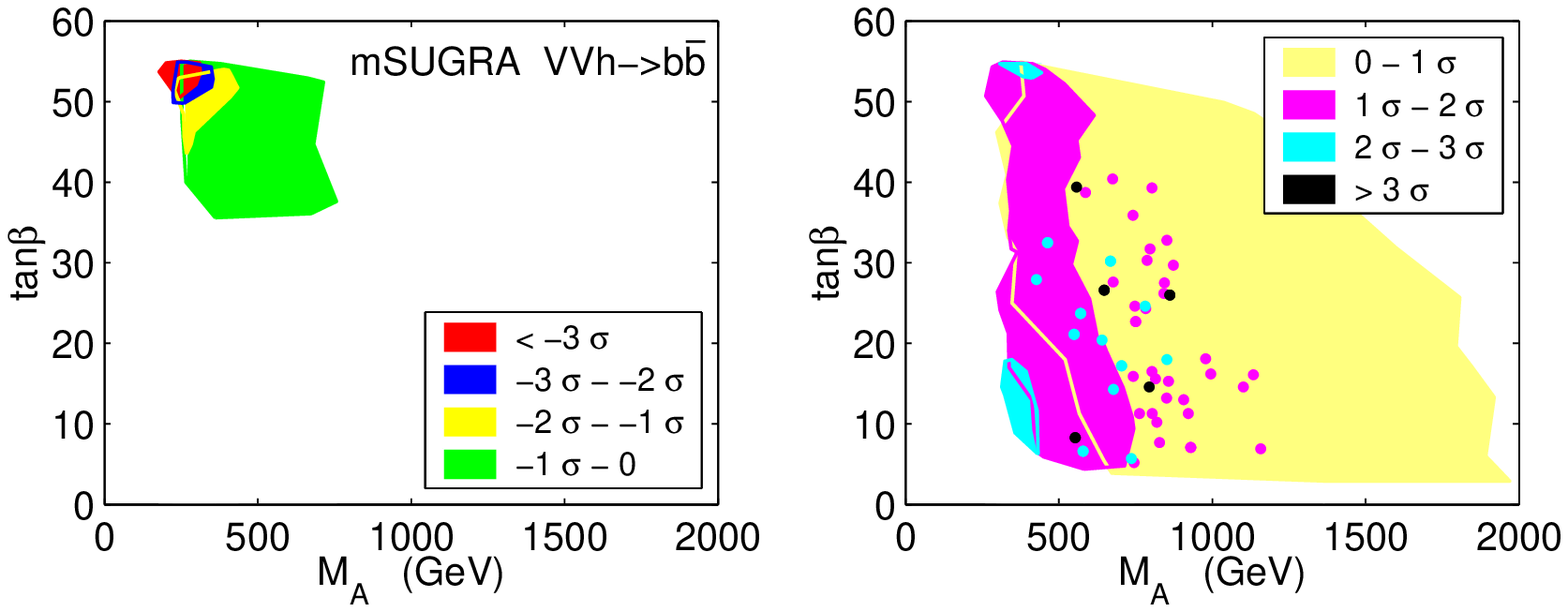,width=12cm,height=5.0cm}
\mbox{}\vspace{1em}
\epsfig{figure=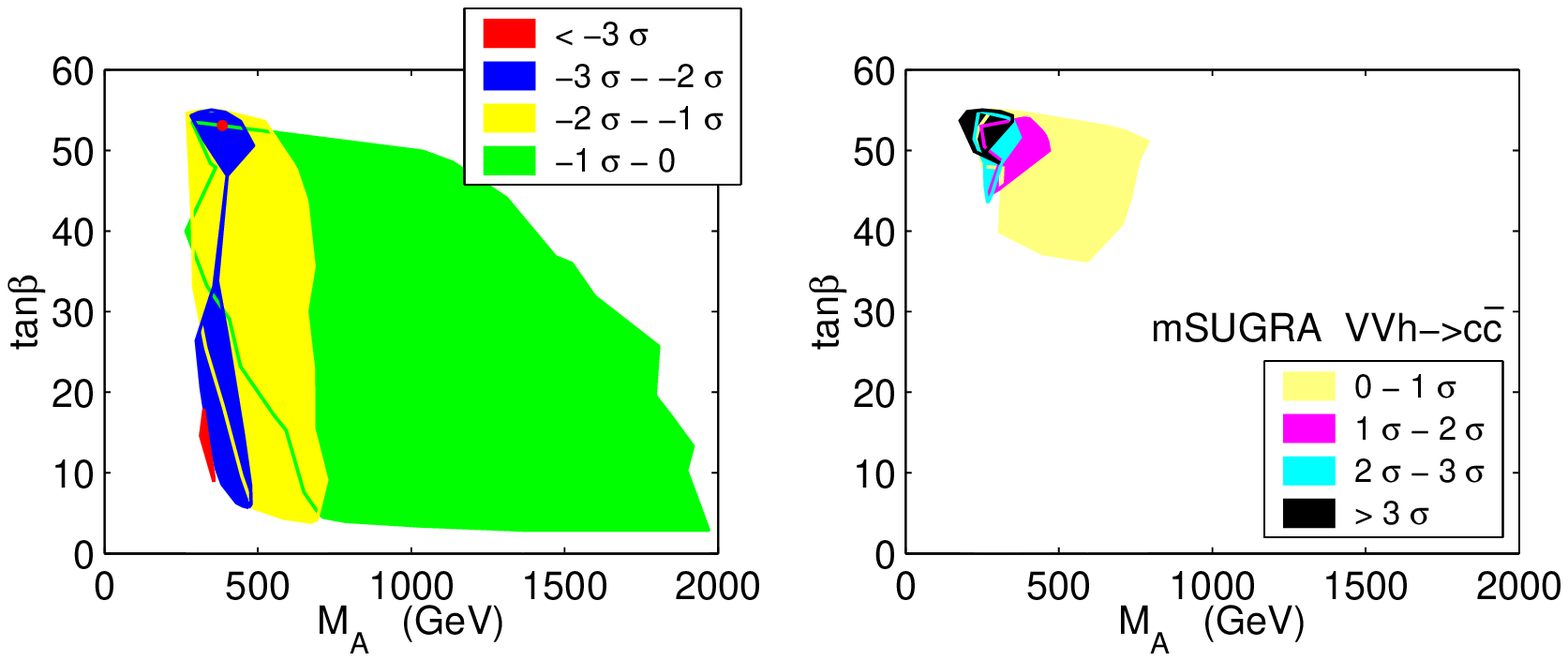,width=12cm,height=5.0cm}
\mbox{}\vspace{1em}
\epsfig{figure=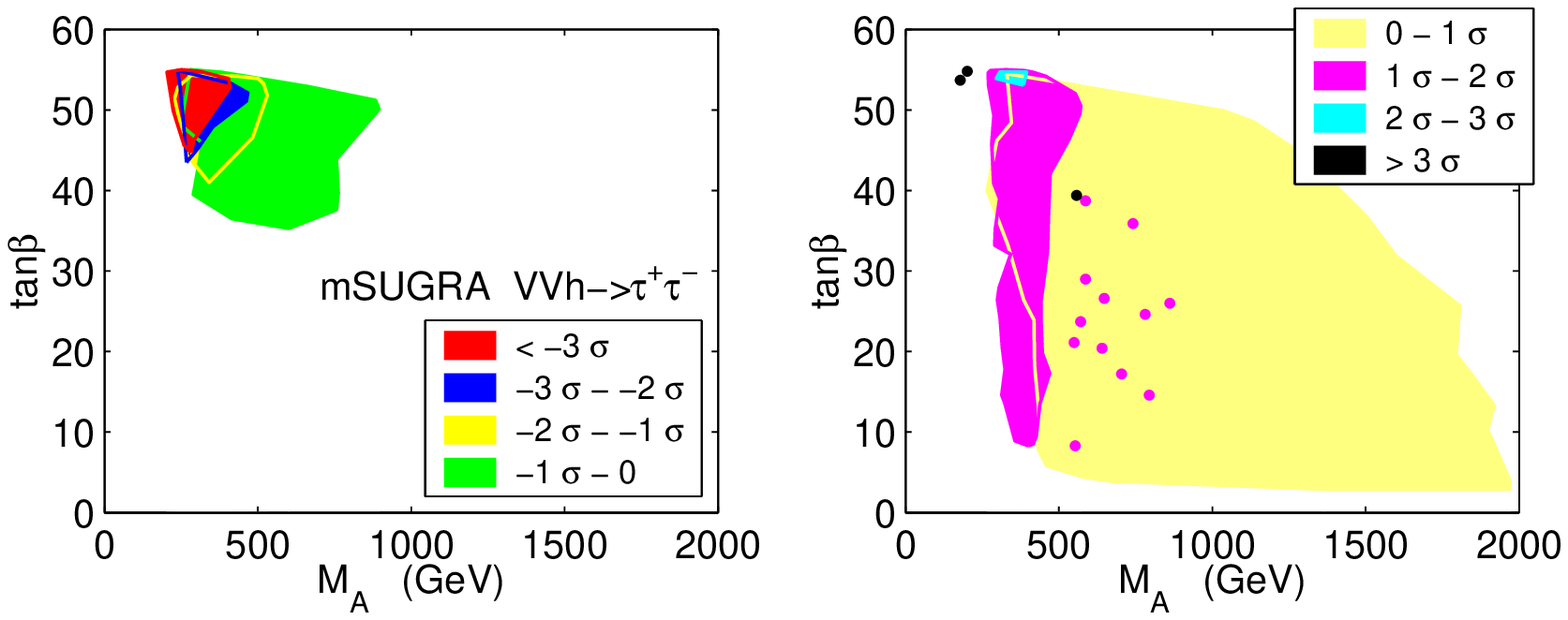,width=12cm,height=5.0cm}
\mbox{}\vspace{1em}
\epsfig{figure=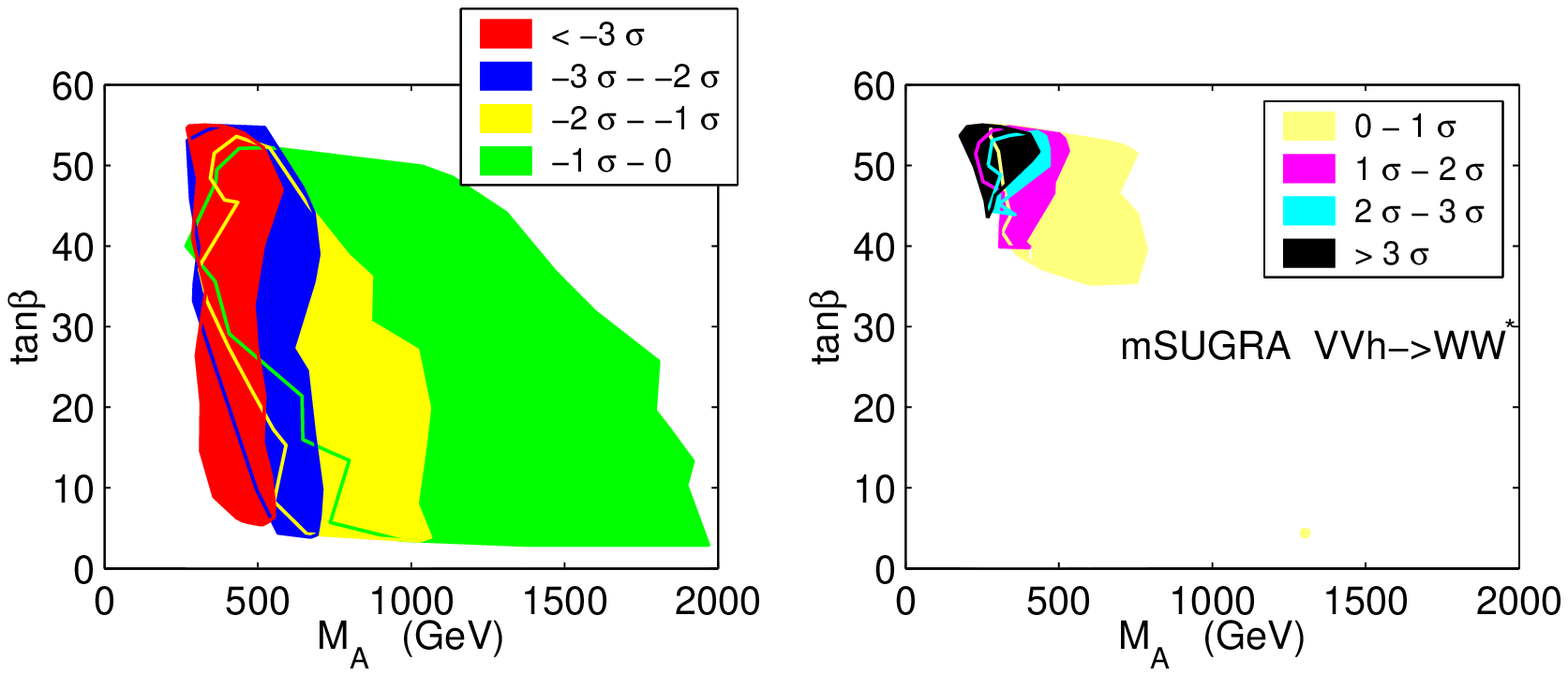,width=12cm,height=5.0cm}
\caption{
Indirect sensitivity to $\MA$ in the mSUGRA scenario: for the channels
$\hbb$, $\hcc$, $\htautau$ and $\hWW$ (from top to bottom) the regions 
in the $\MA$--$\tan\be$ plane are shown where the result in the mSUGRA 
scenario differs from the SM prediction by 1$\si$, 2$\si$ or 3$\si$, 
assuming the prospective accuracy at the LC according to \refta{tab:BRunc}. 
}
\label{mSUGRA_MA_LC}
\end{center}
\vspace{-3em}
\end{figure}
%%%%%%%%%%%%%%%%%%%%%%%%%

\smallskip
In \reffi{mSUGRA_MA_LC} the indirect sensitivity to $\MA$ within the
mSUGRA scenario is investigated for the channels
$h \to b \bar b, c \bar c, \tau^+\tau^-, WW^*$ at the LC. 
The figure shows the regions in the $\MA$--$\tan\be$ plane where the
result in the mSUGRA scenario differs from the SM prediction by 1$\si$,
2$\si$ or 3$\si$, according to the prospective accuracy at the LC as
given in \refta{tab:BRunc}. The corresponding sensitivities at the \gaC\ 
(which are not shown here) turn out to be usually worse than at the LC 
for the mSUGRA scenario. 

If a 2$\si$ or 3$\si$ deviation of the Higgs branching ratios from the
corresponding SM values is found at the LC, an upper bound on $\MA$ can
be inferred within the mSUGRA scenario according to
\reffi{mSUGRA_MA_LC}. In particular, the $\hWW$ channel yields an upper
bound on $\MA$ of $500 - 600 \gev$ (depending on $\tb$) for a more 
than 3$\si$ deviation, $600 - 700 \gev$ for a deviation in excess of 2$\si$,
while deviations of more than 1$\si$ occur for $\MA$ up to $800 - 1000 \gev$
within the mSUGRA scenario.
On the other hand, measuring a suppression in the $\hbb$ and/or
$\htautau$ channel (left column of \reffi{mSUGRA_MA_LC}) or an 
enhancement in the $\hcc$ and/or $\hWW$
channel (right column of \reffi{mSUGRA_MA_LC}) 
would determine $\tb$ to lie within $35 \lsim \tb \lsim 55$ in the
mSUGRA scenario.
The mSUGRA scenario is the only of the three soft SUSY-breaking
scenarios that could accommodate a suppression of the $\hbb$ and/or
$\htautau$ channel%
\footnote{
In our \order{50000} mGMSB scatter points we have found two points with 
$\MA \approx 100 \gev$ and $\tb \approx 55$ that exhibit a very
strong suppression of the $\hbb$ and $\htautau$ channel by more than
50\%. However, these points appear to be rather fine-tuned and we did
not include them into our analysis.
}%
. Thus these measurements can help
to distinguish the soft SUSY-breaking scenarios, see
\refse{subsec:ASBSdisc}.

%%%%%%%%%%%%%%%%%%%%%%%%%%
\begin{figure}[ht!]
\begin{center}
\epsfig{figure=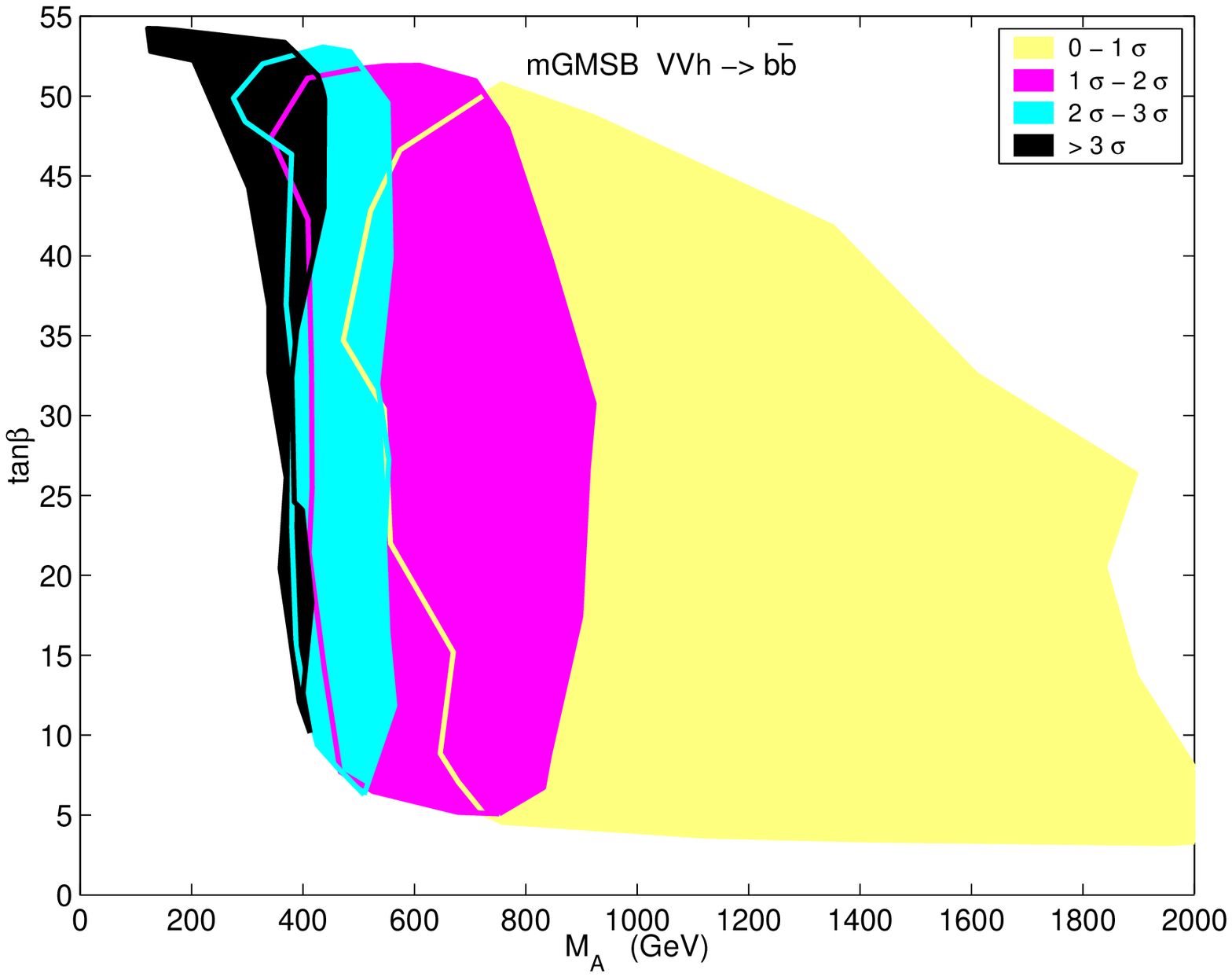,width=6cm,height=5cm}
\epsfig{figure=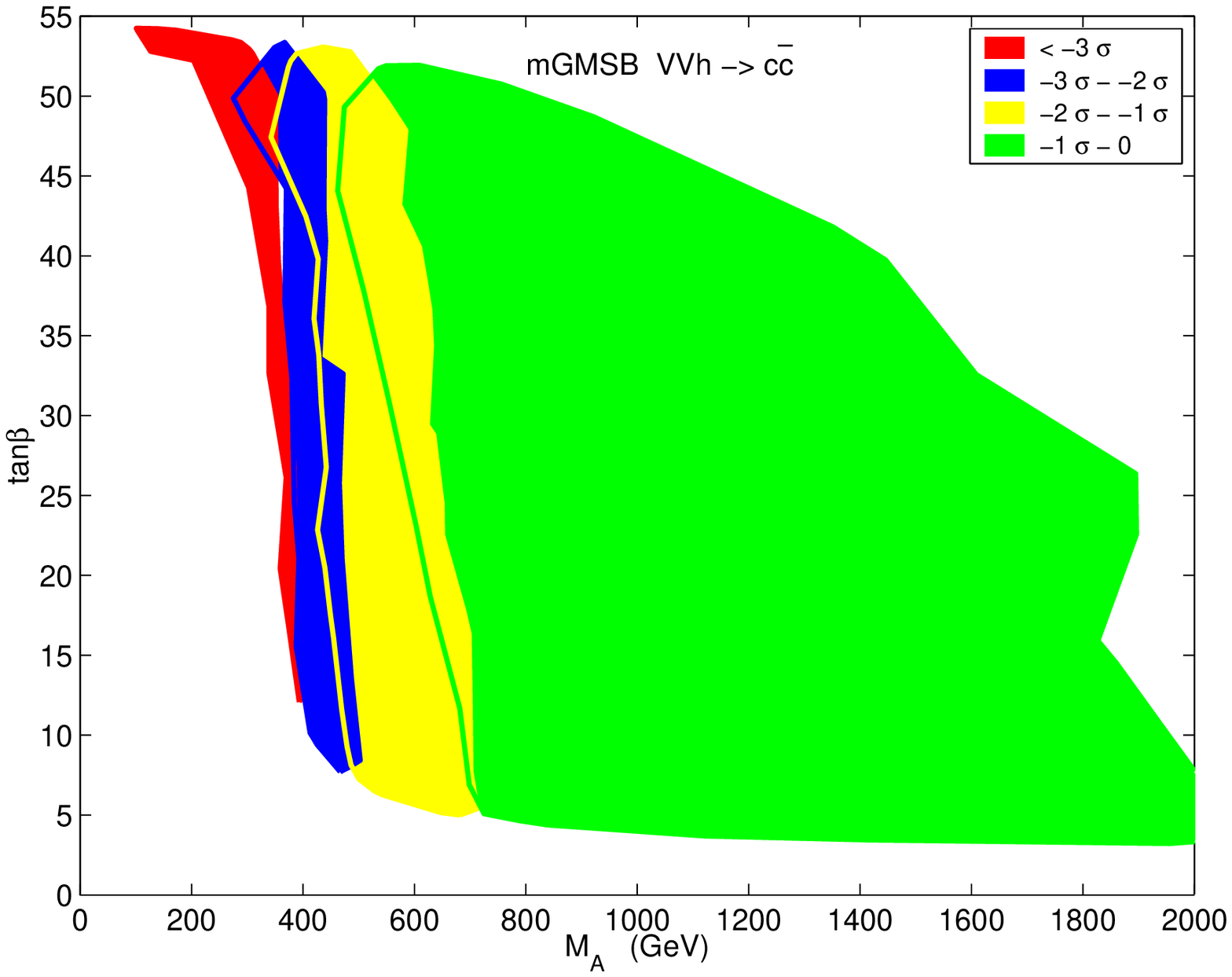,width=6cm,height=5cm}
\mbox{}\vspace{2em}
\epsfig{figure=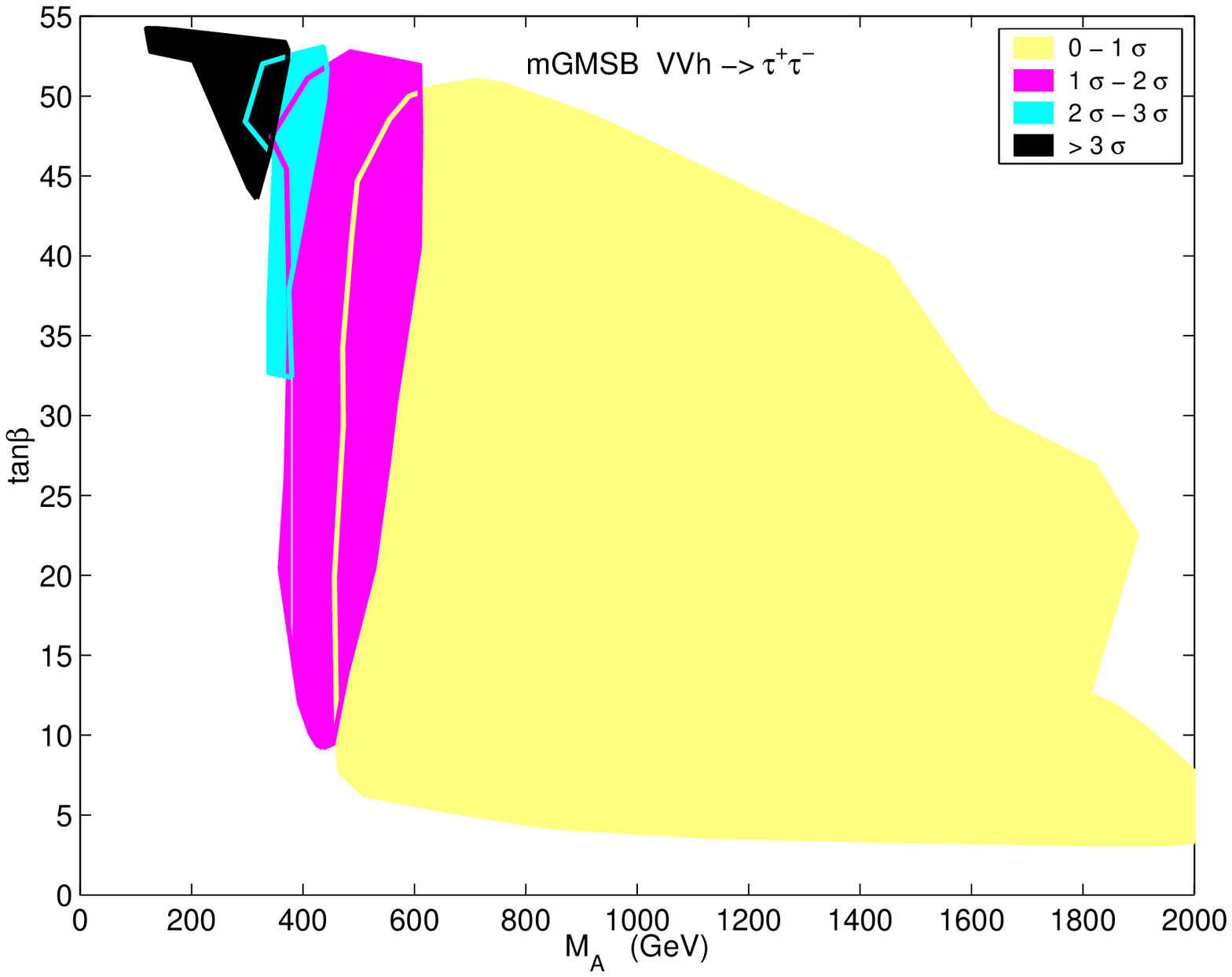,width=6cm,height=5cm}
\epsfig{figure=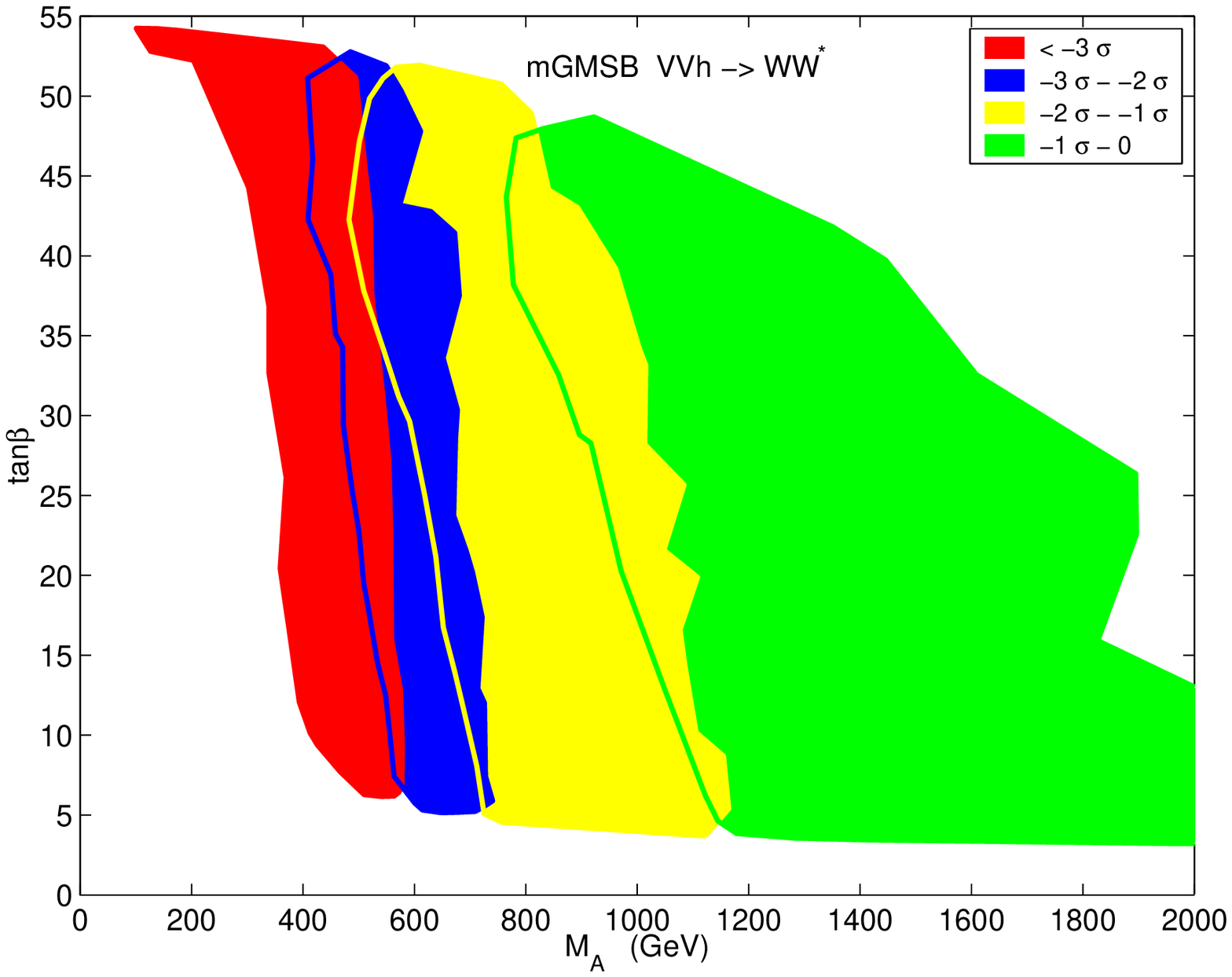,width=6cm,height=5cm}
\vspace{1em}
\caption{
Indirect sensitivity to $\MA$ in the mGMSB scenario: for the channels
$\hbb$ (top left), $\hcc$ (top right), $\htautau$ (bottom left) and
$\hWW$ (bottom right) the regions 
in the $\MA$--$\tan\be$ plane are shown where the result in the mGMSB 
scenario differs from the SM prediction by 1$\si$, 2$\si$ or 3$\si$, 
assuming the prospective accuracy at the LC according to \refta{tab:BRunc}. 
}
\label{GMSB_MA_LC}
\end{center}
\vspace{-1em}
\end{figure}
%%%%%%%%%%%%%%%%%%%%%%%%%

%%%%%%%%%%%%%%%%%%%%%%%%%%
\begin{figure}[ht!]
\vspace{1em}
\begin{center}
\epsfig{figure=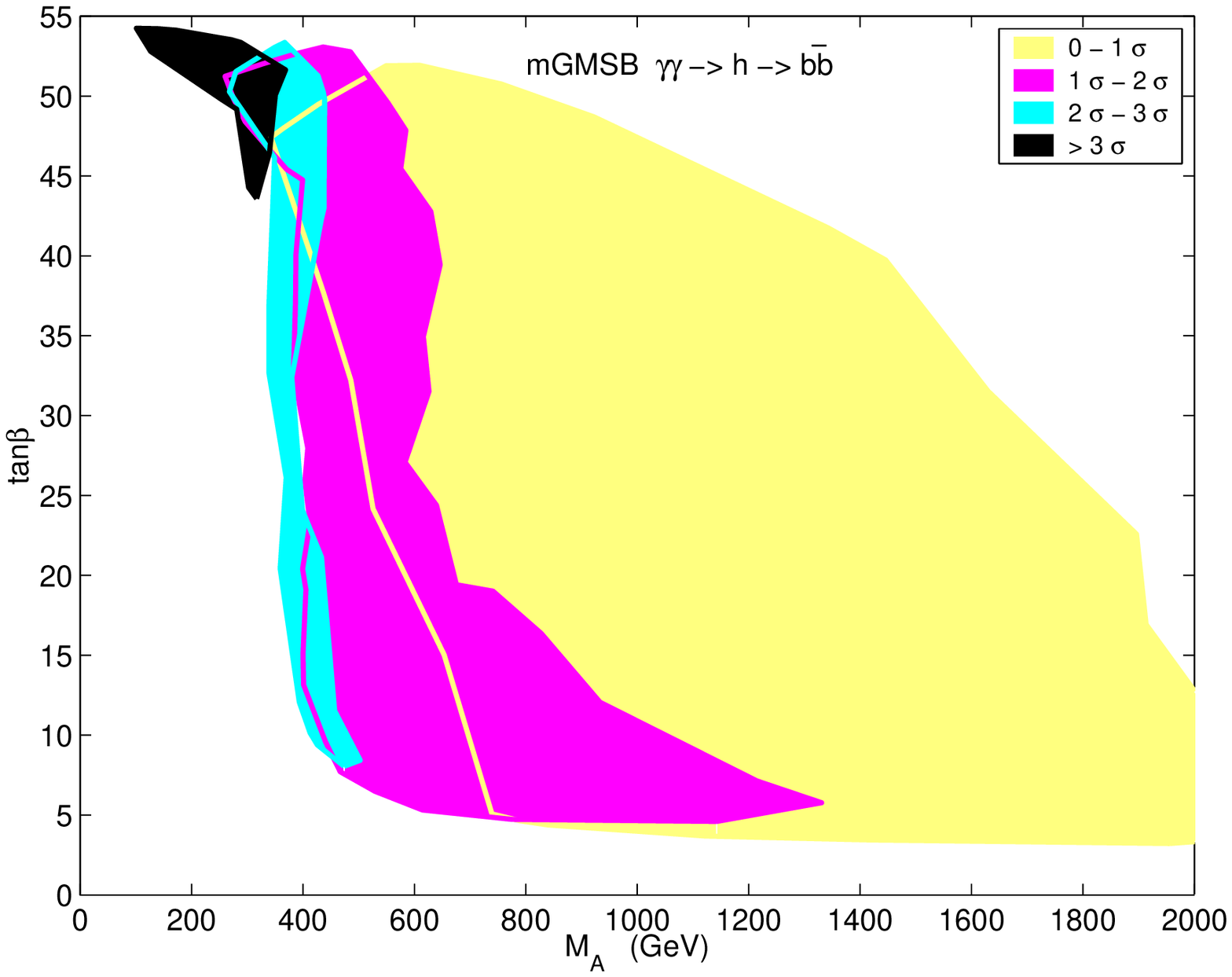,width=6cm,height=5cm}
\epsfig{figure=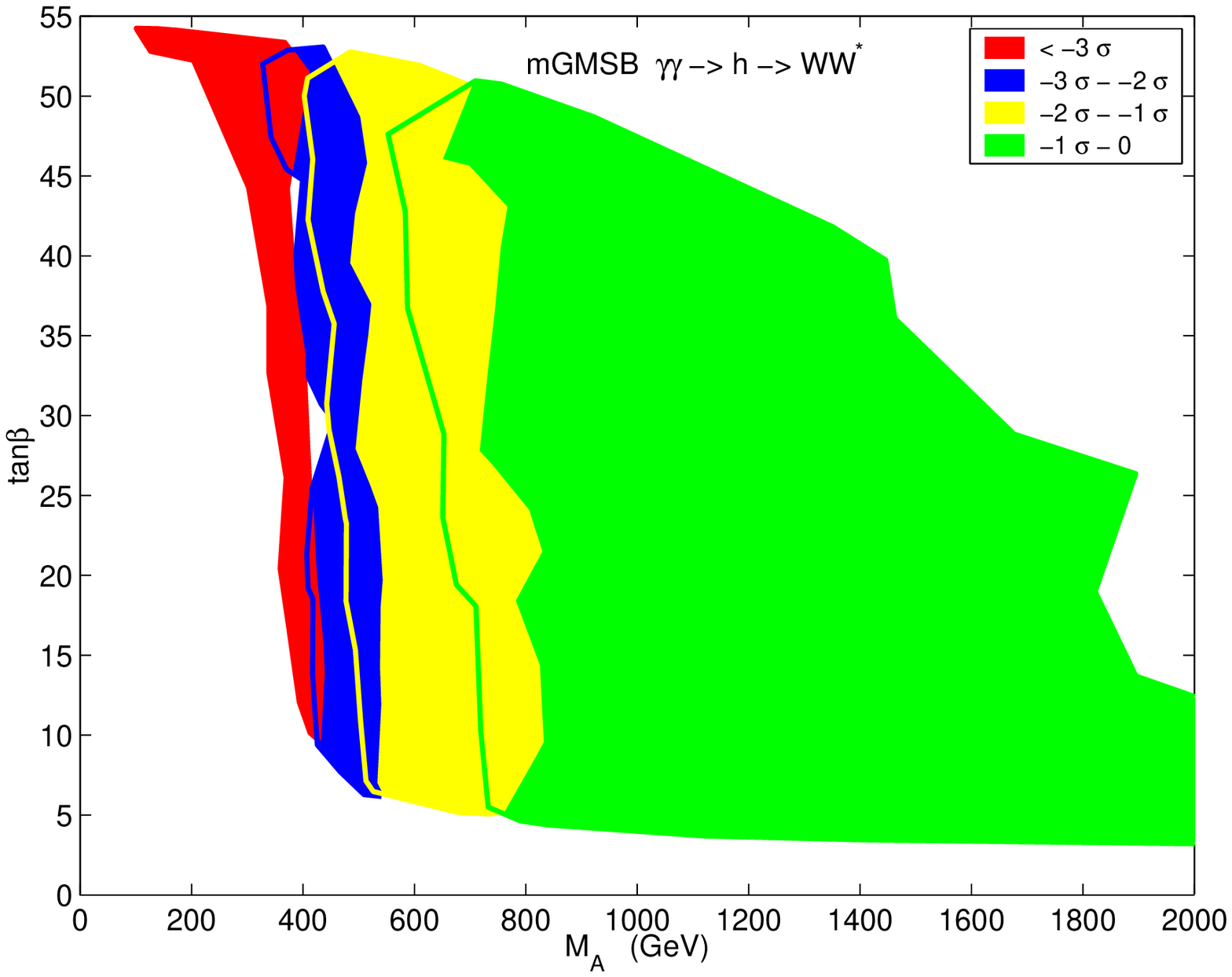,width=6cm,height=5cm}
\vspace{1em}
\caption{
Indirect sensitivity to $\MA$ in the mGMSB scenario: for the channels
$\hbb$ (left) and $\hWW$ (right) the regions
in the $\MA$--$\tan\be$ plane are shown where the result in the mGMSB 
scenario differs from the SM prediction by 1$\si$, 2$\si$ or 3$\si$, 
assuming the prospective accuracy at the \gaC\ according to \refta{tab:BRunc}. 
}
\label{GMSB_MA_gaC}
\end{center}
%\vspace{-1em}
\end{figure}
%%%%%%%%%%%%%%%%%%%%%%%%%
%
%%%%%%%%%%%%%%%%%%%%%%%%%%
\begin{figure}[htb!]
\begin{center}
\epsfig{figure=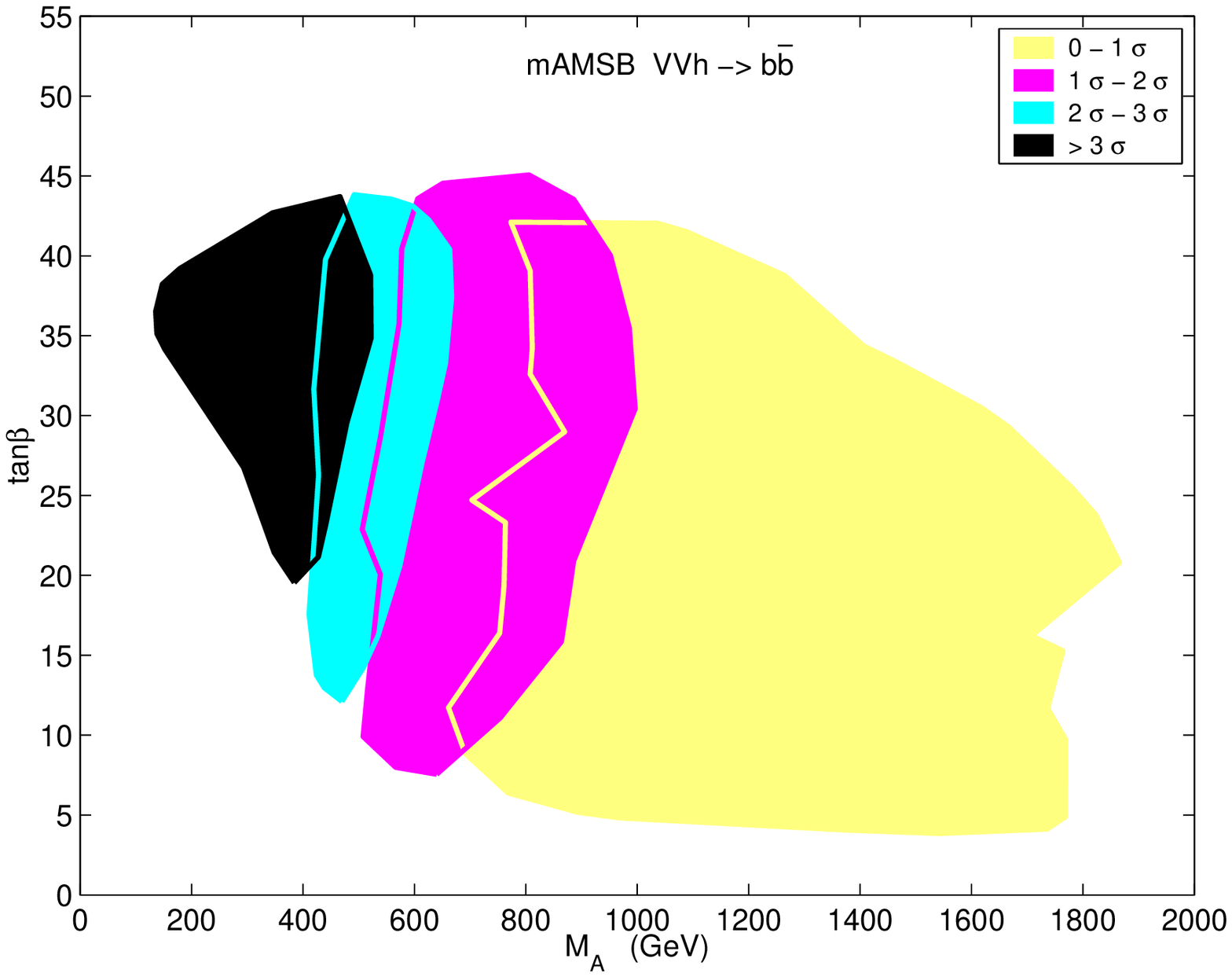,width=6cm,height=5cm}
\epsfig{figure=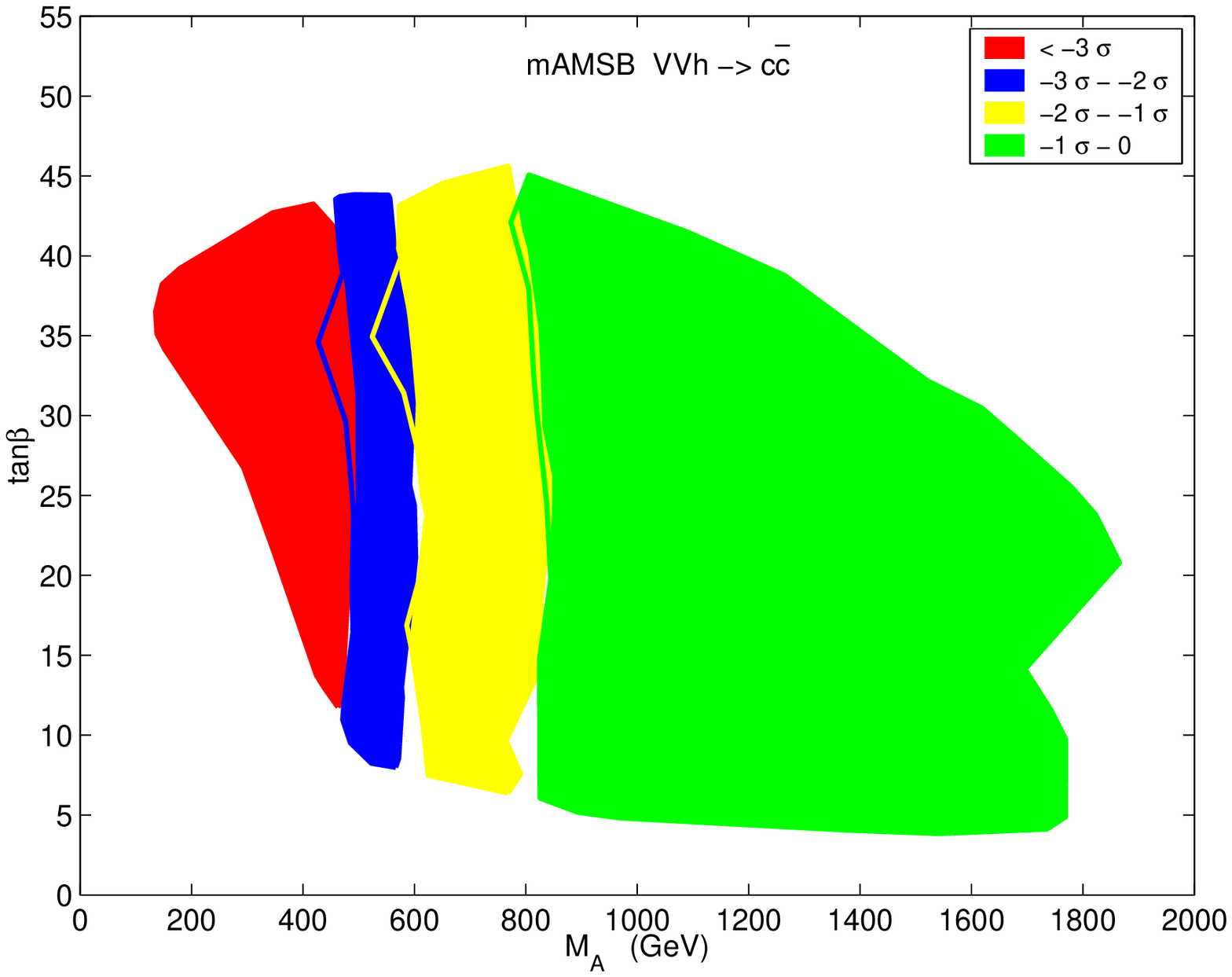,width=6cm,height=5cm}
\mbox{}\vspace{2em}
\epsfig{figure=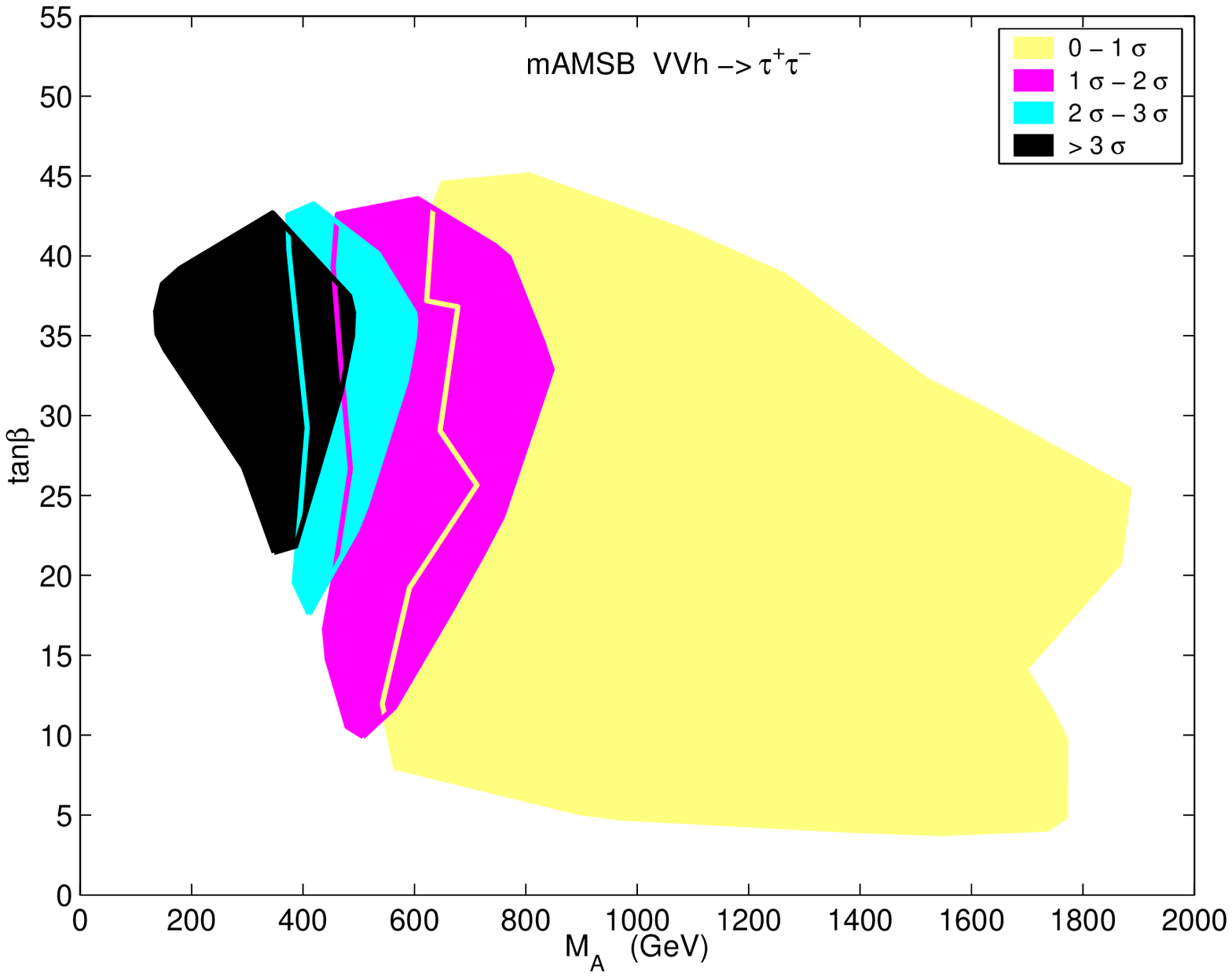,width=6cm,height=5cm}
\epsfig{figure=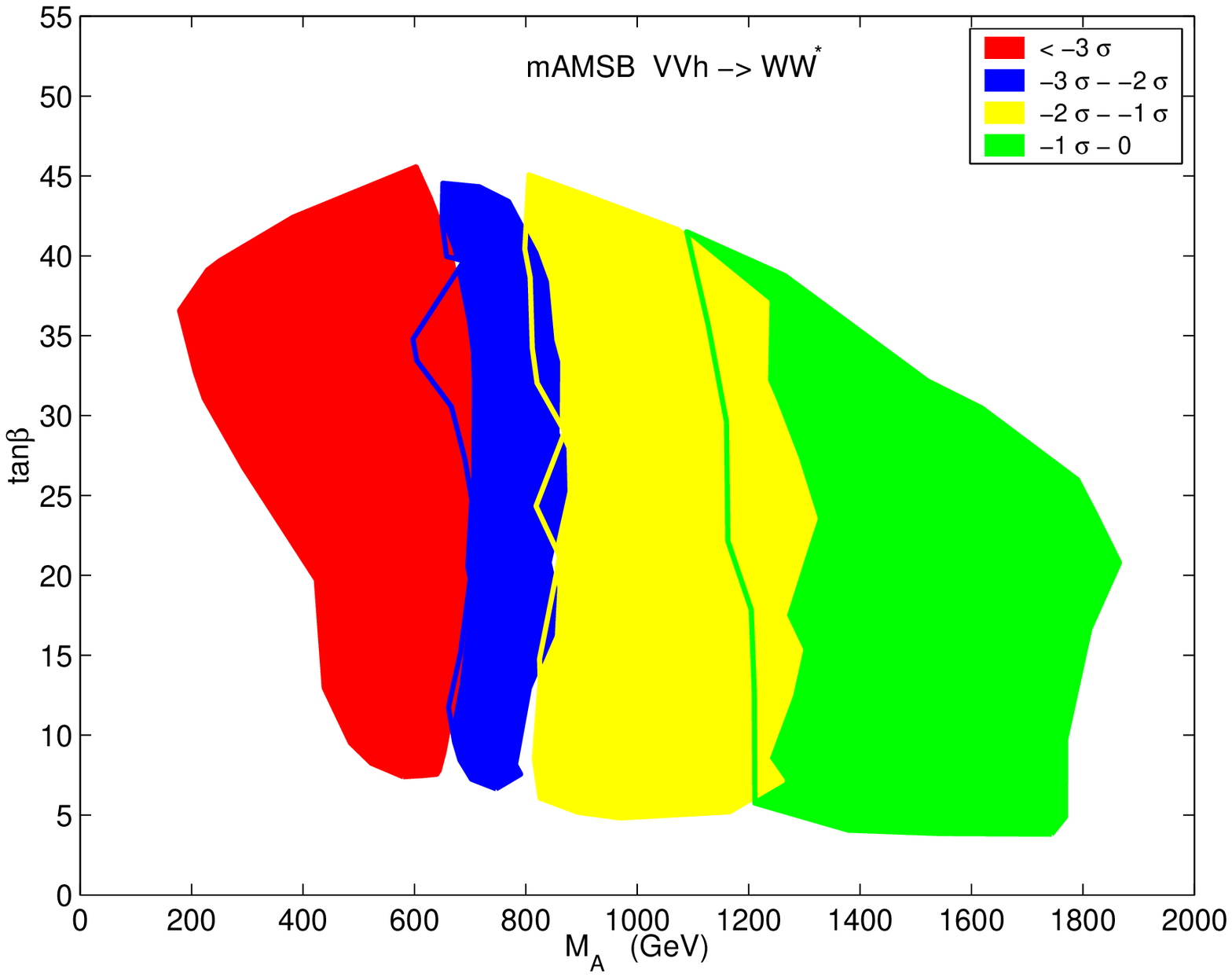,width=6cm,height=5cm}
\vspace{1em}
\caption{
Indirect sensitivity to $\MA$ in the mAMSB scenario: for the channels
$\hbb$ (top left), $\hcc$ (top right), $\htautau$ (bottom left) and
$\hWW$ (bottom right) the regions 
in the $\MA$--$\tan\be$ plane are shown where the result in the mAMSB 
scenario differs from the SM prediction by 1$\si$, 2$\si$ or 3$\si$, 
assuming the prospective accuracy at the LC according to \refta{tab:BRunc}. 
}
\label{AMSB_MA_LC}
\end{center}
\vspace{-4em}
\end{figure}
%%%%%%%%%%%%%%%%%%%%%%%%%

In \reffi{GMSB_MA_LC} we show the sensitivity to $\MA$ within the mGMSB
scenario for the channels 
$h \to b \bar b, c \bar c, \tau^+\tau^-, WW^*$ at the LC. The
corresponding results for $h \to b \bar b, WW^*$ at the \gaC\ are
displayed in \reffi{GMSB_MA_gaC} (which yields comparable sensitivities 
in this scenario).
As for mSUGRA, the observation of a 2$\si$ or 3$\si$ deviation compared
to the SM prediction will allow to establish an upper bound on $\MA$
within the mGMSB scenario. Also in this case the $\hWW$ channel shows
significant deviations from the SM prediction over a wider range of 
the $\MA$--$\tan\beta$ plane than the other channels. For a more than
2$\si$ deviation in this channel an upper bound on $\MA$ of about $700
\gev$ (depending somewhat on $\tb$) can be inferred. 
Bigger deviations result in correspondingly lower upper bounds on $\MA$.

In \reffi{AMSB_MA_LC} the sensitivity to $\MA$ within the mAMSB
scenario is displayed for the channels 
$h \to b \bar b, c \bar c, \tau^+\tau^-, WW^*$ at the LC. The
sensitivities at the \gaC\ are usually worse in this scenario.
As for the other two scenarios, in general an upper bound on $\MA$ can
be established if a 2$\si$ or 3$\si$ deviation from the SM result is
observed. Again in particular the $\hWW$ channel offers good prospects
for observing sizable deviations. It allows to set an upper bound on
$\MA$ of $800 - 900 \gev$ (depending on $\tb$) if a deviation of more
than 2$\si$ is observed.
Higher deviations result in correspondingly lower upper
bounds on $\MA$. Comparing the results for the $\hWW$ channel in the 
mAMSB scenario with the other scenarios, in the mAMSB scenario sizable
deviations from the SM prediction occur over a wider parameter space in
the $\MA$--$\tan\beta$ plane than in the other scenarios. Thus, the
prospects for experimentally establishing a deviation from the SM
prediction and in this way inferring an upper bound on $\MA$ appear to be
particularly good in the mAMSB scenario.

%%%%%%%%%%%%%%%%%%%%%%%%%%%%%%%%%%%%%%%%%%%%%%%%%%%%%%%%%%%%%%
%%%%%%%%%%%%%%%%%%%%%%%%%%%%%%%%%%%%%%%%%%%%%%%%%%%%%%%%%%%%%%

\subsection{Sensitivity to high-energy parameters}
\label{subsec:HEsensitivity}

\begin{figure}[ht!]
\vspace{-1em}
\begin{center}
\epsfig{figure=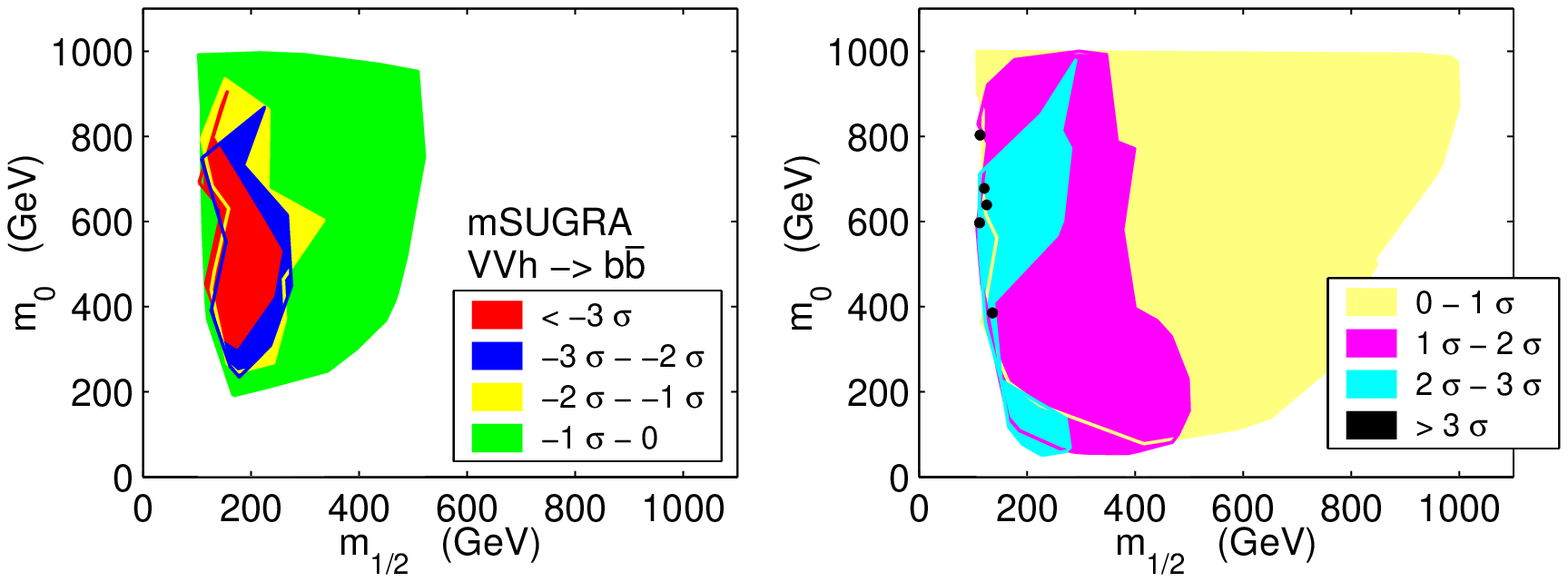,width=12cm,height=5.1cm}
\mbox{}\vspace{1em}
\epsfig{figure=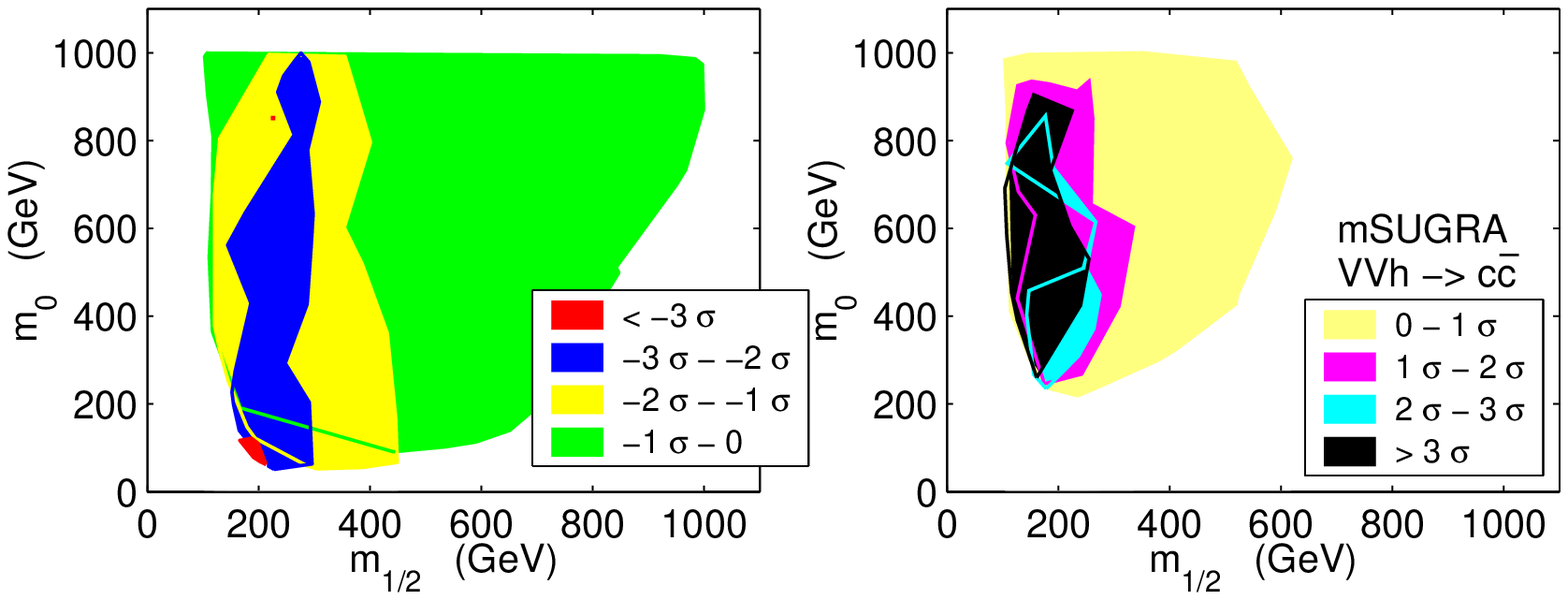,width=12cm,height=5.1cm}
\mbox{}\vspace{1em}
\epsfig{figure=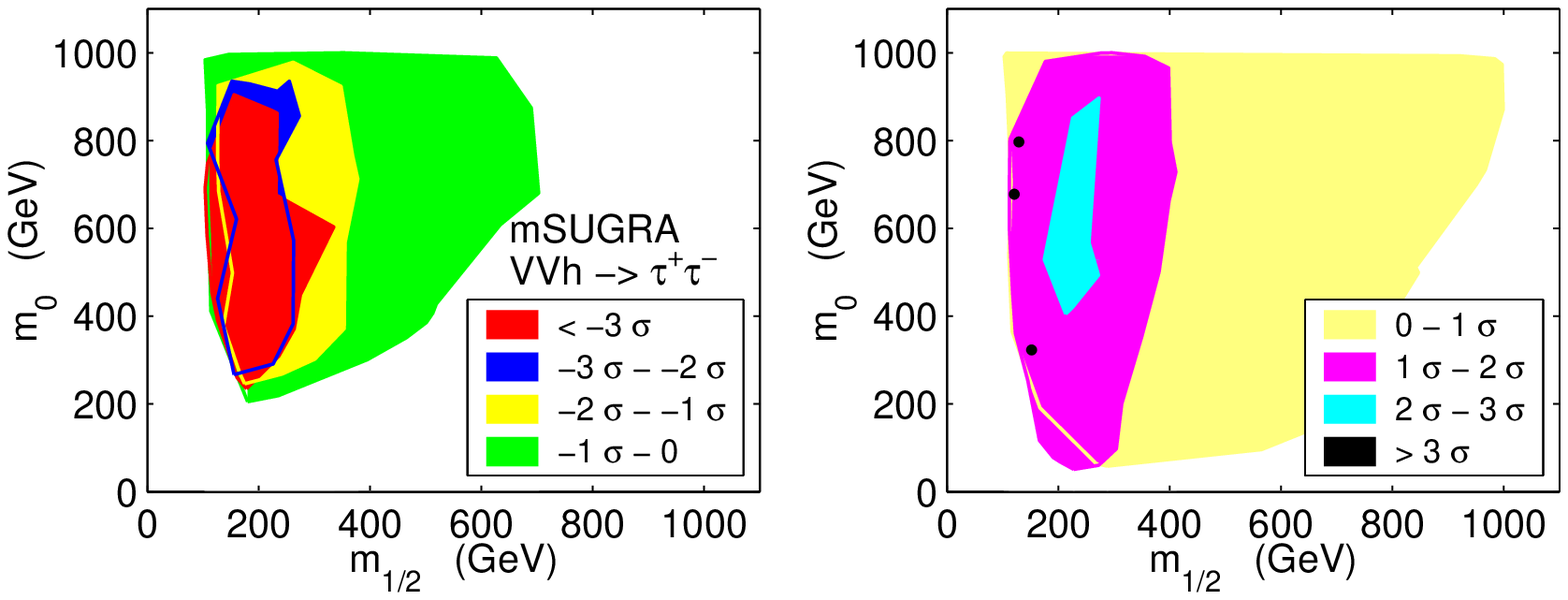,width=12cm,height=5.1cm}
\mbox{}\vspace{1em}
\epsfig{figure=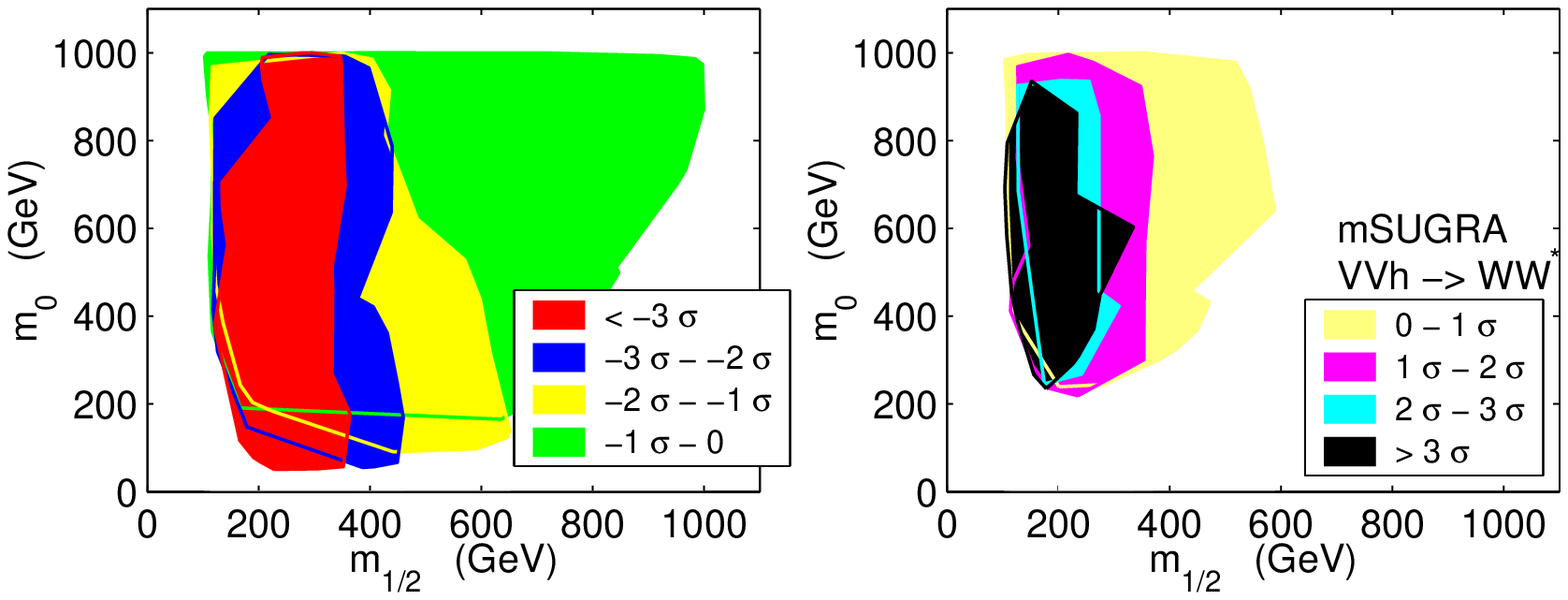,width=12cm,height=5.1cm}
\caption{
Indirect sensitivity to $m_0$, $m_{1/2}$ in the mSUGRA scenario: for the
channels $\hbb$, $\hcc$, $\htautau$ and $\hWW$ (from top to bottom)
the regions in the $m_{1/2}$--$m_0$ plane are shown where the result in
the mSUGRA scenario differs from the SM prediction by 1$\si$, 2$\si$ or 
3$\si$, assuming the prospective accuracy at the LC according to
\refta{tab:BRunc}.
}
\label{mSUGRA_GUT_LC}
\end{center}
\vspace{-4em}
\end{figure}

Besides providing sensitivity to $\MA$,
precise measurements of Higgs branching rations at the LC can also
yield indirect information on the high-energy parameters of the
different soft SUSY-breaking scenarios.
In \reffi{mSUGRA_GUT_LC} the results for the channels
$h \to b \bar b, c \bar c, \tau^+\tau^-, WW^*$ are shown in the
$m_{1/2}$--$m_0$ plane for the mSUGRA scenario. 
While the indirect constraints that can be obtained with a
2$\si$ or 3$\si$ deviation on $m_0$ are rather mild, stronger bounds
can be obtained for $m_{1/2}$. 
This reflects the fact that $\MA$ and the squark masses are strongly
correlated with the $m_{1/2}$ value. Combining the channels, an
upper bound of $\sim 350 \gev$ on $m_{1/2}$ can be set if a deviation 
of more than 3$\si$ from the SM prediction is observed, a 2$\si$
deviation constrains $m_{1/2}$ to be smaller than $\sim 450 \gev$,
while deviations of more than 1$\si$ occur for $m_{1/2} \lsim 650 \gev$.

%%%%%%%%%%%%%%%%%%%%%%%%%%
%%%%%%%%%%%%%%%%%%%%%%%%%

Concerning the mGMSB scenario (which is not displayed here) the
indirect constraints are weaker. Results deviating from the SM
prediction for the $\hWW$ channel by 3$\si$, for instance, are
distributed over nearly the whole $M_{\rm mess}$--$\La$ plane. Thus,
establishing a non SM-like behavior in the Higgs sector alone is not
sufficient to derive indirect bounds on $M_{\rm mess}$ and $\La$, 
further experimental information is necessary to constrain these
parameters. On the other hand, weak {\em lower} limits on $M_{\rm
mess}$, $\La$ could be set, which can cut out the lower edge of the
mGMSB allowed $M_{\rm mess}$--$\La$ area, if the deviation from the SM
value is found to be small.

%%%%%%%%%%%%%%%%%%%%%%%%%%
\begin{figure}[htb!]
\begin{center}
\epsfig{figure=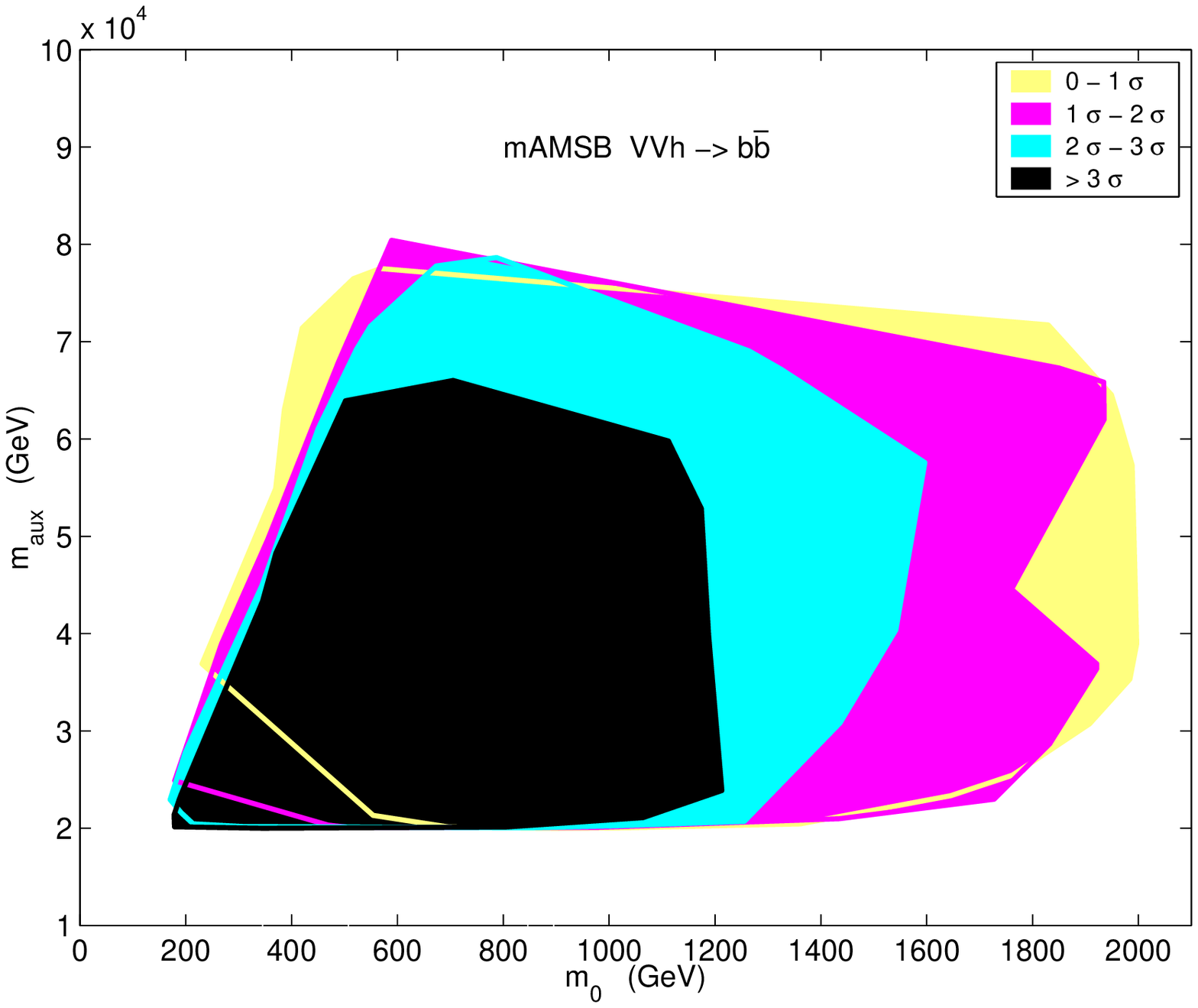,width=6cm,height=5cm}
\epsfig{figure=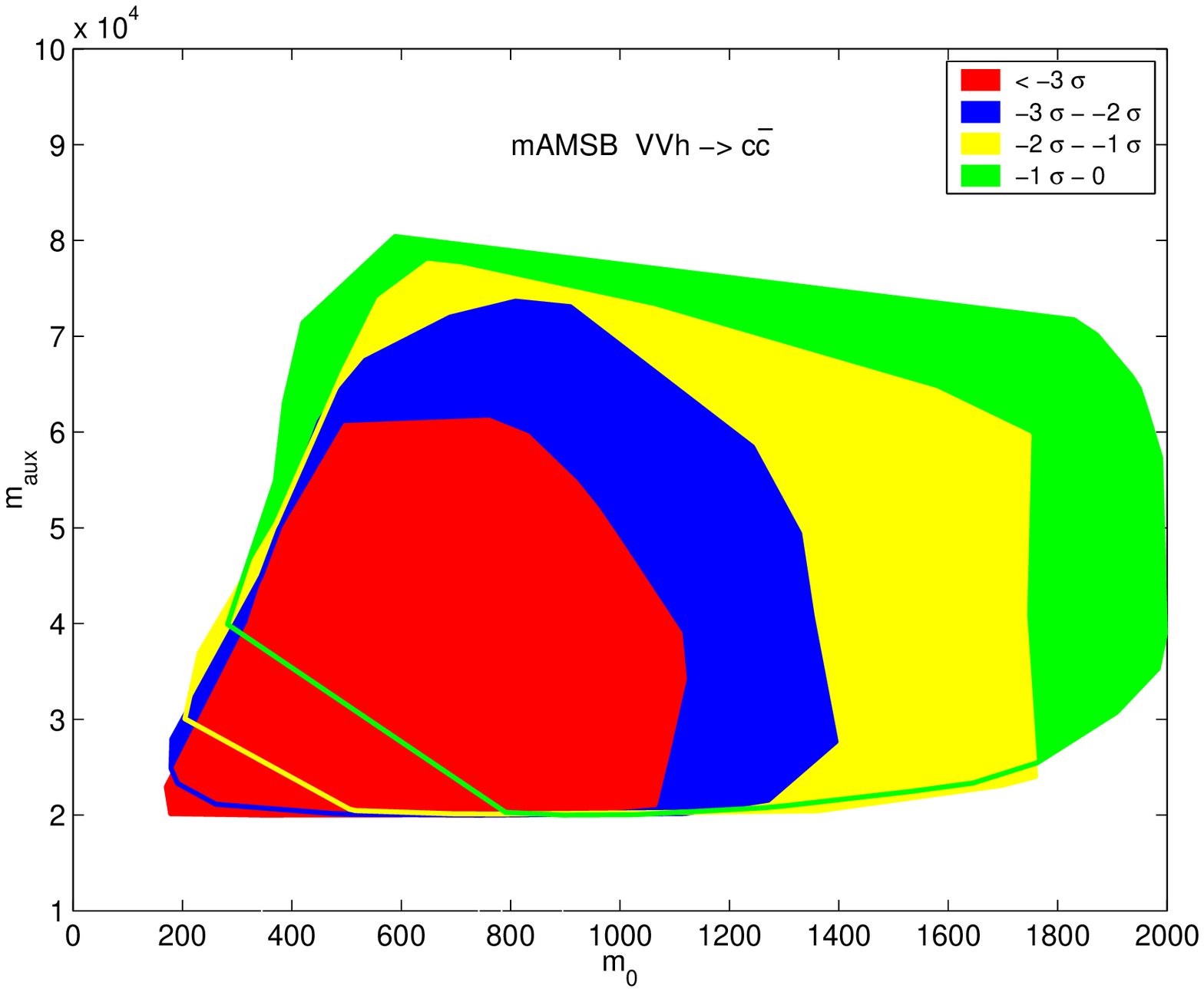,width=6cm,height=5cm}
\caption{
Indirect sensitivity to $m_0$, $m_{\rm aux}$ in the mAMSB scenario: for 
the channels $\hbb$ (left) and $\hcc$ (right) the regions in the 
$m_0$--$m_{\rm aux}$ plane are shown where the result in
the mAMSB scenario differs from the SM prediction by 1$\si$, 2$\si$ or 
3$\si$, assuming the prospective accuracy at the LC according to
\refta{tab:BRunc}.
}
\label{AMSB_GUT_LC}
\end{center}
\vspace{-1em}
\end{figure}
%%%%%%%%%%%%%%%%%%%%%%%%%

In \reffi{AMSB_GUT_LC} the results for the channels
$h \to b \bar b, c \bar c$ are shown in the
$m_0$--$m_{\rm aux}$ plane for the mAMSB scenario. 
Like in the mGMSB scenario, deviations of 3$\si$ or 2$\si$ with respect to
the SM prediction occur over a rather wide range of $m_0$ and $m_{\rm aux}$ 
values. An observed deviation of 3$\si$ would constrain $m_0$ to be
smaller than $\sim 1100 \gev$, while $m_{\rm aux}$ would have to be
smaller than $\sim 6 \cdot 10^4 \gev$. Observation of a 2$\si$ deviation
would allow to set an upper bound on $m_0$ of $m_0 \lsim 1400 \gev$,
while restricting the parameter space to the one compatible with a 
2$\si$ deviation does not significantly reduce the range of possible values 
of $m_{\rm aux}$ in \reffi{AMSB_GUT_LC}.

%%%%%%%%%%%%%%%%%%%%%%%%%%%%%%%%%%%%%%%%%%%%%%%%%%%%%%%%%%%%%%
%%%%%%%%%%%%%%%%%%%%%%%%%%%%%%%%%%%%%%%%%%%%%%%%%%%%%%%%%%%%%%

\subsection{Discrimination between soft SUSY-breaking scenarios}
\label{subsec:ASBSdisc}

We now investigate the potential of precise measurements of Higgs branching
ratios at a LC for distinguishing between the three soft SUSY-breaking
scenarios. The main experimental test of different SUSY-breaking
scenarios will of course be to confront their predictions for the SUSY
spectrum with the results obtained from the direct observation of SUSY
particles. The analysis of the Higgs sector, however, may contribute
further valuable information. Since the different soft SUSY-breaking
scenarios predict different mass patterns for the SUSY particles,
variations in the results for the Higgs sector observables for the
same value of $\MA$ can be expected.

As starting point of our analysis, 
as above, we do not assume experimental input from other sectors of the
MSSM, but concentrate on the Higgs sector. Clearly, resolving
differences between the predictions of the three scenarios via
measurements in the Higgs sector will require some experimental
information on the tree-level parameters of the Higgs sector, $\MA$ and
$\tb$. Therefore we will focus on a scenario where the LHC can detect
the heavy MSSM Higgs bosons via their decays $H/A \to \tau^+\tau^-$
(with the main production channel $b \bar b \to b \bar b \; H/A$), which
can be realized for large $\tb$ and not too large 
$\MA$~\cite{atlastdr,LHHiggs2001}. As a specific example we assume that
the LHC provides a measurement of $\MA$ as well as a lower bound on 
$\tb$, 
\BE
500 \gev \lsim \MA \lsim 600 \gev, \quad \tb \gsim 30~.
\label{tbMAatLHC}
\EE
The results of the analysis below would improve for smaller values of
$\MA$, while for a larger $\MA$ and smaller $\tb$ observation of the
heavy Higgs bosons at the LHC would become increasingly difficult.
Restricting the data set of our scan, see \refse{subsec:susybreak}, to
those parameter points fulfilling \refeq{tbMAatLHC} we compare the
predictions for the different branching ratios arising from the three
scenarios. As above, we indicate the deviations from the SM prediction
in terms of the prospective accuracy at the LC according to
\refta{tab:BRunc}.

%%%%%%%%%%%%%%%%%%%%%%%%%%
\begin{figure}[ht!]
\vspace{2cm}
\begin{center}
\epsfig{figure=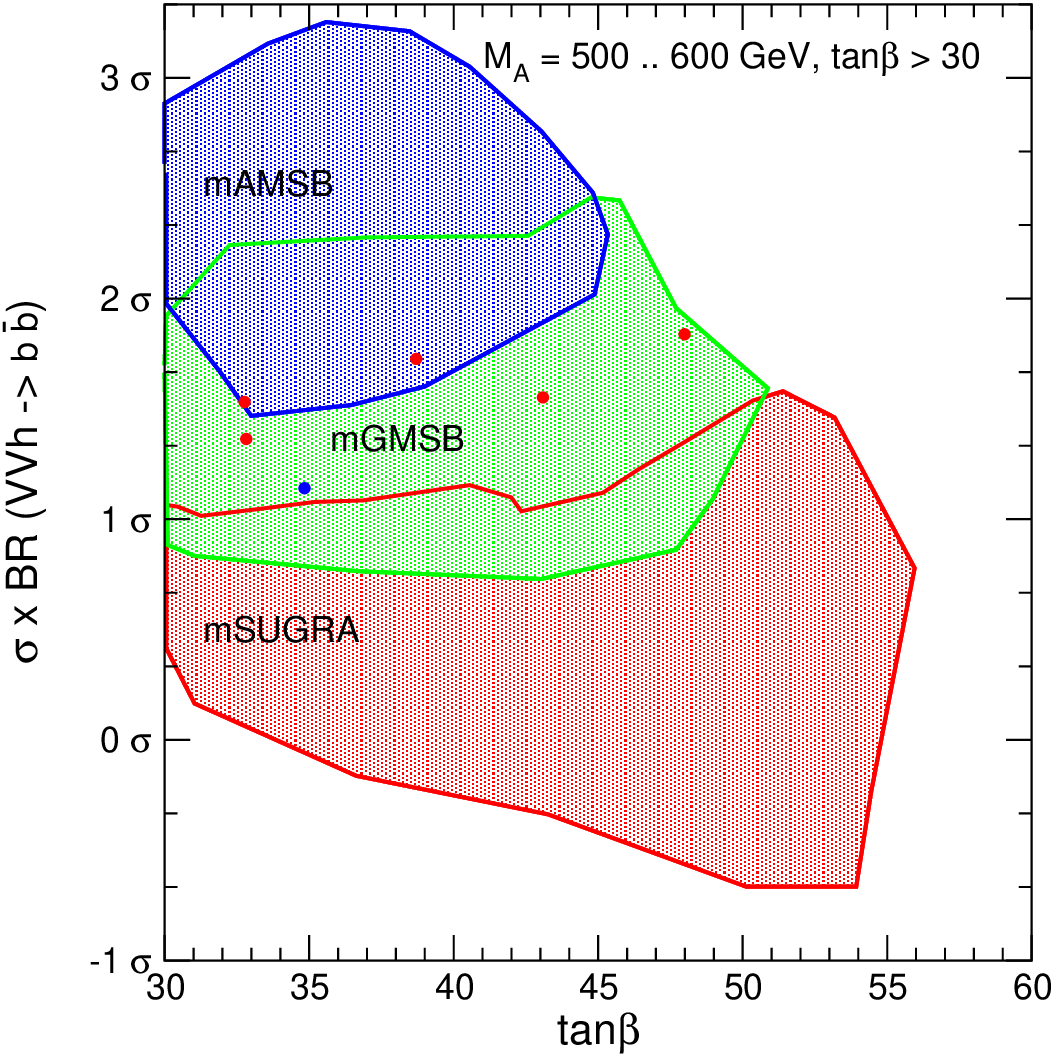,width=12cm,height=8cm}
\mbox{}\vspace{3em}
\epsfig{figure=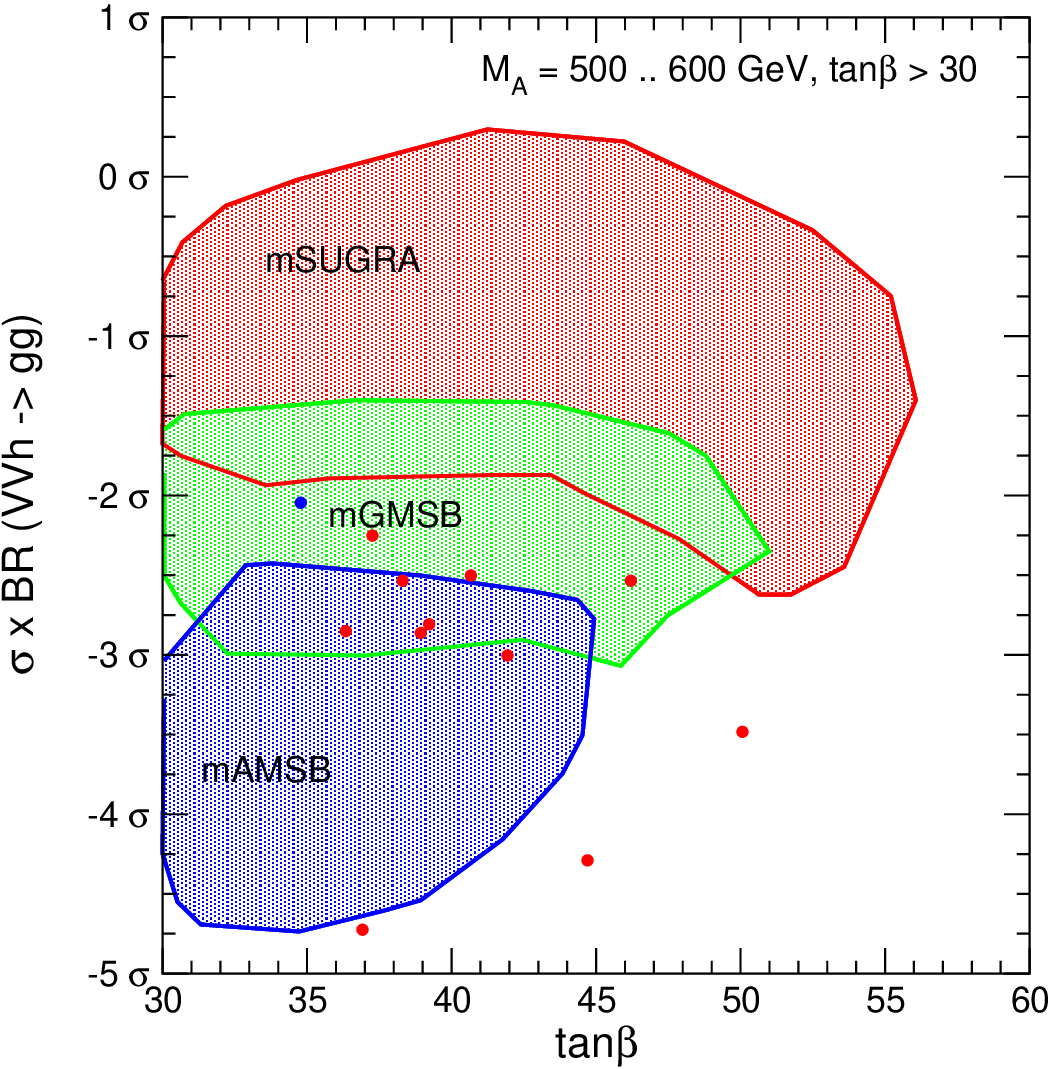,width=12cm,height=8cm}
\vspace{1cm}
\caption{
Comparison of $\br(\hbb)$ (top) and $\br(\hgg)$ (bottom) in the three
soft SUSY-breaking scenarios via LC measurements.
}
\label{ASBS_BR}
\end{center}
\end{figure}
%%%%%%%%%%%%%%%%%%%%%%%%%

In \reffi{ASBS_BR} we show the results for the channels $\hbb$ and
$\hgg$. The results for these channels, as for the others that are not
shown ($\htautau$, $\hcc$ and $\hWW$), are similar and show the following
general pattern (see also the discussion in \refse{sec:higgsobs} and 
\refse{subsec:MAsensitivity}): 
for the $\MA$ values
corresponding to \refeq{tbMAatLHC} the mAMSB scenario gives rise to 
larger deviations in the branching ratios from the SM values than the
mGMSB scenario. Thus, if in the situation of \refeq{tbMAatLHC} a $3\si$
deviation from the SM value were found in $\br(\hbb)$ and a $-4\si$
deviation in $\br(\hgg)$, this would be better compatible
with an AMSB scenario than with a mGMSB scenario. If, on the other hand, 
the branching ratios were found to agree well with the SM prediction,
this would be best compatible with a SUSY-breaking scenario of mSUGRA
type. 

As a consequence, precision measurements at the LC of the branching ratios 
of the light $\cp$-even Higgs boson of the MSSM may indicate a
preference among the three soft SUSY-breaking scenarios at the 1-2$\si$
level. This information will be complementary to the information from
the direct observation of SUSY particles.

The different behavior as a function of $\MA$ in the three scenarios can
be traced back mainly to different loop contributions to the
off-diagonal entry in the Higgs propagator matrix, $\hSi_{\PePz}$, which
according to \refeq{aeff} give rise to differences in the effective
mixing angle $\aeff$ entering the Higgs couplings. 
Especially the dominant decay channel $\hbb$, being $\sim \sin^2\aeff/\CQb$, 
is strongly affected. While in the mSUGRA
scenario $\hSi_{\PePz}$ has in general fairly large and negative values,
in the mGMSB scenario $\hSi_{\PePz}$ is small, and in the mAMSB scenario
it gets large and positive values. In combination with the tree-level
dependence on $\MA$, see \refeq{aeff}, this leads to a different degree
of decoupling with respect to the SM result as function of $\MA$.

So far we have not assumed any additional 
experimental input on the SUSY spectrum
from the Tevatron or the LHC. We have
checked, however, that the results in \reffi{ASBS_BR} are essentially
unmodified if parameter points for which the Tevatron will detect SUSY
particles are excluded from the scan. 

%%%%%%%%%%%%%%%%%%%%%%%%%%
\begin{figure}[ht!]
\vspace{1cm}
\begin{center}
\epsfig{figure=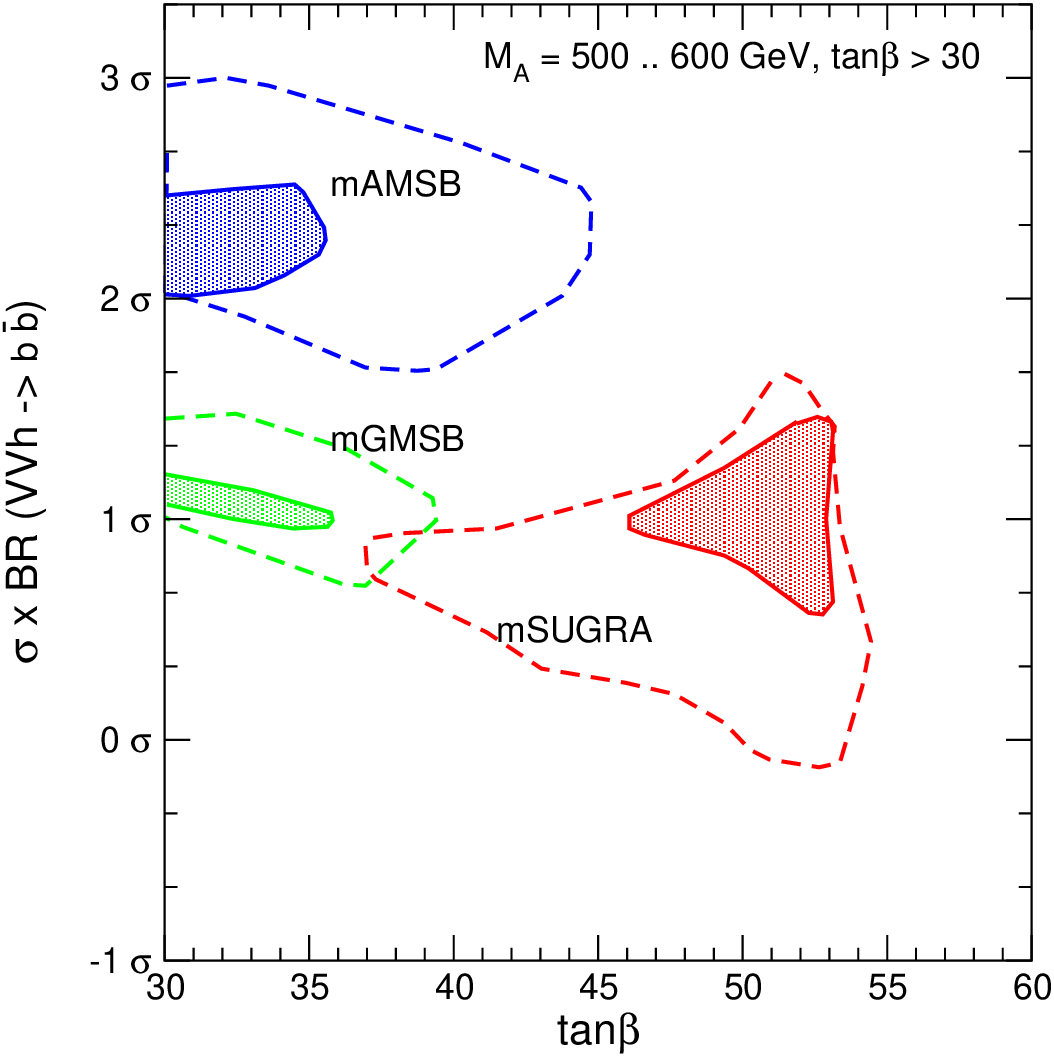,width=12cm,height=8cm}
\mbox{}\vspace{1em}
\epsfig{figure=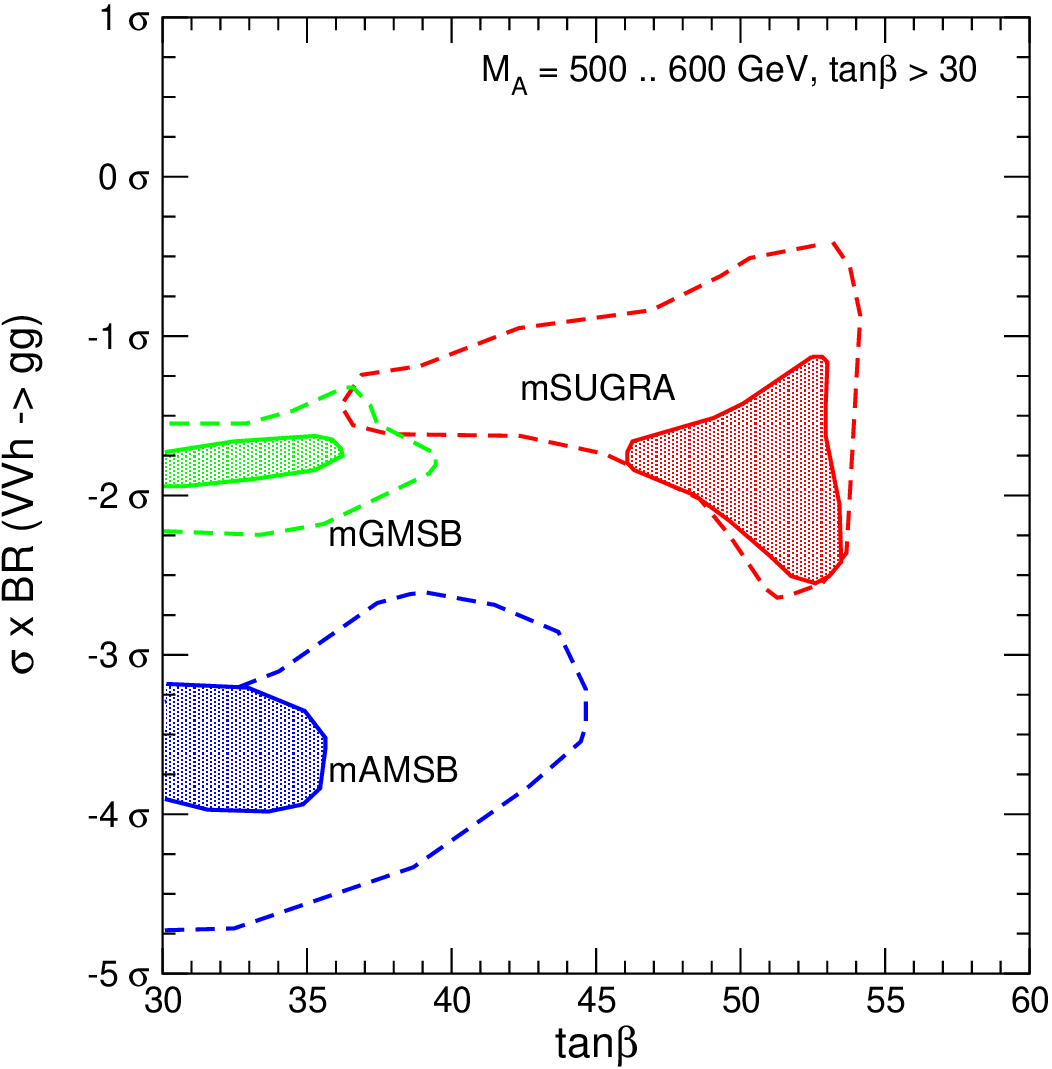,width=12cm,height=8cm}
\vspace{.8cm}
\caption{
Comparison of $\br(\hbb)$ (top) and $\br(\hgg)$ (bottom) in the three
soft SUSY-breaking scenarios via LC measurements, assuming direct input
on the SUSY spectrum from the LHC. The areas surrounded by dashed lines
correspond to the parameter regions in the three scenarios where the
light scalar top mass lies in the region 
$800 \gev \leq \mste \leq 900 \gev$, 
while the shaded areas surrounded by full lines correspond
to the case where furthermore the gluino mass is known to be
constrained by $900 \gev \leq \mgl \leq 1000 \gev$. 
\label{ASBS_BR_LHC}
}
\end{center}
\end{figure}
%%%%%%%%%%%%%%%%%%%%%%%%%

Concerning possible experimental
information on the SUSY spectrum from the LHC, the situation strongly
depends on the assumed scenario. For illustration we thus restrict to
one particular example, shown in \reffi{ASBS_BR_LHC}.
The regions indicated by dashed lines 
correspond to parameter regions in the three scenarios where
experimental information on the light scalar top quark is assumed,
\begin{equation}
800 \gev \lsim \mste \lsim 900 \gev .
\end{equation}
The shaded areas surrounded by full lines correspond
to the case where furthermore the gluino mass is assumed to be bounded
by
\begin{equation}
900 \gev \lsim \mgl \lsim 1000 \gev . 
\end{equation}
As expected, assuming direct experimental information on the SUSY
spectrum in addition to measurements in the Higgs sector significantly
enhances the sensitivity for distinguishing between the different soft
SUSY-breaking scenarios.
While for the particular scenario studied here it is not possible to
distinguish between the mGMSB and mSUGRA scenarios on the basis of the
Higgs branching ratios alone, additional information on $\tb$ would
allow a clear distinction.

%%%%%%%%%%%%%%%%%%%%%%%%%%%%%%%%%%%%%%%%%%%%%%%%%%%%%%%%%%%%%%
%%%%%%%%%%%%%%%%%%%%%%%%%%%%%%%%%%%%%%%%%%%%%%%%%%%%%%%%%%%%%%

\section{Conclusions}
\label{sec:conclusions}

We have investigated the relevant production and decay channels of the
lightest $\cp$-even MSSM Higgs boson at the Tevatron, the LHC, 
an $e^+e^-$~LC, a \gaC\ and a \muC\ within the mSUGRA, mGMSB
and mAMSB scenarios. The values of 
$\si \times \br$ have been compared with the corresponding SM values
with the same Higgs boson mass, $\MHSM = \mh$. 
In this context we have also updated earlier results on the upper bound
on $\mh$ within the three scenarios and on the lower bounds on $\tb$
that can be inferred by confronting the theoretical predictions with 
the LEP exclusion limit.

We have first analyzed the observability of the lightest MSSM Higgs
boson at the different colliders. 
The modes $gg \to h \to \ga\ga$, $t\bar t \to t\bar t h$ and 
$WW \to h \to WW^*, \tau^+\tau^-, \ga\ga$ allow the detection of the
lightest MSSM 
Higgs boson in all three scenarios over the whole corresponding
parameter space.
Possible exceptions occur for the very small $\MA$ region in the
mGMSB and mAMSB scenarios, 
where a strong suppression of more than 50\% could happen for
$gg \to h \to \ga\ga$.
Within the clean 
experimental environment of the LC the observation of the light Higgs will 
be ensured for all three scenarios. For a \gaC, the very small $\MA$
region in mGMSB and mAMSB can be problematic for the $\hbb$ and
$\htautau$ mode. At the \muC, $\MA \lsim 150 \gev$ and $\tb \gsim 50$
for mSUGRA exhibits a strong suppression for the $\hbb$ and $\htautau$ 
mode, while on the other hand in this parameter region the production of
the heavy MSSM Higgs bosons $H$, $A$ happens with an enhanced rate.
Besides these difficult regions, the main search modes at a \gaC\ and
a \muC\ do not suffer from severe suppressions with respect to the SM
case in all three scenarios. The results of this analysis are summarized
in \refta{tab:summary}. Thus, all possible future colliders offer
very good prospects for detecting the lightest $\cp$-even Higgs boson
of the MSSM in an mSUGRA, mGMSB or mAMSB scenario.

We then investigated the potential of precision measurements of Higgs 
branching ratios at the LC and the \gaC\ for establishing indirect
constraints on $\MA$ and $\tb$. For this analysis we have not assumed
any further experimental information on the SUSY spectrum, i.e.\ a full
scan over the parameter space (restricting to the case $\mu > 0$) has
been performed. If deviations of the Higgs branching ratios from their
SM values will be found at the 2--$3\si$ level, it will be possible to
establish an upper bound for $\MA$ significantly below 1~TeV in all
three scenarios. The biggest sensitivity will come from the $\hWW$ and
$\hcc$ channels. Within the mSUGRA scenario, furthermore a bound on
$\tb$ of $35 \lsim \tb \lsim 55$ can be obtained if a suppression of 
the $\hbb$ and/or $\htautau$ channel or an enhancement in the $\hcc$ 
and/or $\hWW$ channel with respect to the SM values is observed.
If this would be the case, this could be independently confirmed by
Higgs mediated B-physics observables like $B^0\to \mu^+\mu^-$ or
$B^0-\bar{B^0}$ mixing.

Similarly, precise measurements of $\si \times \br$ at the LC can also 
provide indirect information on the high-energy parameters of the three
soft SUSY-breaking scenarios. While within the mGMSB scenario the
experimental determination of the Higgs branching ratios will allow to
set only very weak bounds on the high-energy parameters, within mSUGRA
relatively strong bounds on $m_{1/2}$ and in mAMSB moderate bounds on
$m_0$ could be set.

Finally we have investigated the potential of precise measurements of
$\si \times \br$ at a LC to distinguish between the three soft
SUSY-breaking scenarios. For this analysis we have assumed a situation
where experimental information on $\MA$ (and to a lesser extent on
$\tb$) obtained at the LHC 
can be combined with precision measurements of the properties of
the light Higgs boson at the LC, see also \citere{lhclc}. If a
significant suppression of the $\hbb$ and/or $\htautau$ channel with
respect to its SM value were found, this would point towards the mSUGRA
scenario, irrespectively of the actual value of $\MA$ 
(with $\MA \lsim 1$~TeV). Otherwise, assuming in our example $\MA$ to be
restricted to $500 \gev \lsim \MA \lsim 600 \gev$, precise measurements
of $\si \times \br$ in particular in the $\hbb$ and $\hgg$ channels 
may indicate a preference among the three soft SUSY-breaking scenarios
at the 1--2$\si$ level. This information might be valuable as it
complements the one about the SUSY spectrum 
from the direct observation of SUSY particles.

%%%%%%%%%%%%%%%%%%%%%%%%%%%%%%%%%%%%%%%%%%%%%%%%%%%%%%%%%%%%%%
%%%%%%%%%%%%%%%%%%%%%%%%%%%%%%%%%%%%%%%%%%%%%%%%%%%%%%%%%%%%%%

\section*{Acknowledgements}
We thank S.~Ambrosanio for helpful discussion on the GMSB scenario as
well as for his Fortran code {\em SUSYFIRE}.
We thank T.~Plehn for discussions.
A.D.\ thanks P.~Slavich for interesting discussions.
S.S.\ has been supported by the DOE grant DE-FG03-92-ER-40701. 
A.D.\ would like to acknowledge financial support from the
Network RTN European Program HPRN-CT-2000-00148
``Physics Across the Present Energy Frontier: Probing the Origin of
Mass''.
This work has been supported by the European Community's Human
Potential Programme under contract HPRN-CT-2000-00149 Physics at
Colliders.

%%%%%%%%%%%%%%%%%%%%%%%%%%%%%%%%%%%%%%%%%%%%%%%%%%%%%%%%%%%%%%
%%%%%%%%%%%%%%%%%%%%%%%%%%%%%%%%%%%%%%%%%%%%%%%%%%%%%%%%%%%%%%

%\clearpage
%\newpage

%%%%%%%%%%%%%%%%%%%%%%%%%%%%%%%%%%%%%%%%%%%%%%%%%%%%%%%%%%%%%%
%%%%%%%%%%%%%%%%%%%%%%%%%%%%%%%%%%%%%%%%%%%%%%%%%%%%%%%%%%%%%%

\end{document}